\documentclass[12pt]{report}
\usepackage[utf8]{inputenc}
\usepackage[slovak,english]{babel}
\usepackage[a4paper,left=2.5cm,right=2.5cm,top=2.5cm,bottom=2.5cm]{geometry}

\usepackage{amsmath}
\usepackage{amssymb}
\usepackage{tabularx}
\usepackage{graphicx}
\graphicspath{ {./images/} }
\usepackage{caption}
\usepackage{subcaption}
\usepackage{pdfpages}
\usepackage{multirow}
\usepackage{multicol}
\usepackage[colorlinks=true,urlcolor=blue]{hyperref}
\usepackage[normalem]{ulem}
\usepackage[nottoc]{tocbibind}

\usepackage{afterpage}

\def\Gammac{{\stackrel{\circ}{\Gamma}}{}}
\def\Gammab{{\stackrel{}{\Gamma}}{}}
\def\Gammag{{\mathbf{\Gamma}}{}}
\def\Tg{{\mathbf{T}}{}}
\def\Rg{{\mathbf{R}}{}}
\def\Qg{{\mathbf{Q}}{}}

\def\omegag{{\boldsymbol{\omega}}{}}
\def\nablag{{\boldsymbol{\nabla}}{}}
\def\nablac{{\stackrel{\circ}{\nabla}}{}}
\def\Tb{{\stackrel{}{T}}{}}
\def\Rc{{\stackrel{\circ}{R}}{}}
\def\Kc{{\stackrel{\circ}{K}}{}}
\def\omegac{{\stackrel{\circ}{\omega}}{}}
\def\omegab{{\stackrel{}{\omega}}{}}
\def\Kb{{\stackrel{}{K}}{}}
\def\Lc{{\stackrel{\circ}{\mathcal{L}}}{}}
\def\Lb{{\stackrel{}{\mathcal{L}}}{}}
\def\Jb{{\stackrel{}{\mathcal{J}}}{}}
\def\Sb{{\stackrel{}{S}}{}}
\def\tb{{\stackrel{}{t}}{}}
\def\Ssb{{\stackrel{}{\mathcal{S}}}{}}
\def\Ssc{{\stackrel{\circ}{\mathcal{S}}}{}}

\usepackage[sort&compress,numbers]{natbib}

\begin{document}

%thesis_info
\newpage
\thispagestyle{empty}
\addtocounter{page}{-1}
\begin{center}
{\large COMENIUS UNIVERSITY IN BRATISLAVA \\
FACULTY OF MATHEMATICS, PHYSICS AND INFORMATICS}
\end{center}

\vspace{5cm}
\begin{center}
{\large \bf BLACK HOLE ACTION IN EINSTEIN'S OTHER GRAVITY \\
\vspace{3cm}
MASTER THESIS}
\end{center}

\vfill
\begin{flushleft}
\begin{tabular}{ll}
Study programme: & Theoretical Physics \\
Field of study: & Physics \\
Research department: & Department of Theoretical Physics \\
Supervisor: & Mgr. Martin Krššák, Dr.rer.nat \\
\end{tabular}
\end{flushleft}

\vfill
\begin{multicols}{2}
\begin{flushleft} Bratislava 2024 \end{flushleft}
\begin{flushright} {\bf Bc. Michal Stano} \end{flushright}
\end{multicols}

%\afterpage{\blankpage}

%acknowledgements
\newpage
\thispagestyle{empty}
\paragraph*{Acknowledgments}

I would like to express my deepest gratitude towards my supervisor for showing me what it entails to become a physicist. I would also like to give thanks to my parents and to my partner, Lujza, for the constant support, regardless of my somewhat pessimistic worldview.\\ 

\begin{flushright}
    \textit{''When the going gets tough, the tough get going."}
\end{flushright}
\begin{flushright}
    - J. Kennedy
\end{flushright}

\vfill
\paragraph*{Declaration}

I hereby declare that I have written this master's thesis by myself under the guidance of my supervisor and using the literature listed in the bibliography.

\begin{multicols}{2}
{\bf
\begin{flushleft} Bratislava 2024 \end{flushleft}
\begin{flushright} {\bf Bc. Michal Stano} \end{flushright} 
}
\end{multicols}

\newpage
%abstract
\thispagestyle{empty}
\section*{Abstract}

STANO, Michal: Black Hole Action in Einstein's other Gravity [Master thesis], Comenius University in Bratislava, Faculty of Mathematics, Physics and Informatics, Department of Theoretical Physics ; supervisor: Mgr. Martin Krššák, Dr.rer.nat, Bratislava, 2024.\\

The main goal of this thesis is to study the gravitational action for the Schwarzschild black hole and its subsequent regularisation from the perspective of general relativity and teleparallel gravity. The standard approach of general relativity requires the Einstein-Hilbert action to be supplemented with the Gibbons-Hawking-York (GHY) boundary term whenever the underlying spacetime contains a boundary. However, the total action is typically divergent, and thus it needs to be regularised by the so-called background subtraction. Next, we consider the teleparallel geometry characterised by the curvature-free teleparallel spin connection, all the while adopting the tetrad formalism.  In contrast to general relativity, teleparallel spin connection is however left  undetermined by the field equations. We argue that the correct choice of such a connection in principle regularizes the action as equally as the background subtraction, without the need of introducing any additional regulating terms. In the interior region of a black hole, the teleparallel approach agrees completely with the Gibbons-Hawking (GH) method, but in the exterior region, there appears to be a discrepancy in the overall factor of two. We investigate the origin of this problem by testing various frames, including the so-called proper and canonical frames. Finally, we explore the role of singular points in the action and their contributions to the surface terms structure as a possible remedy to this problem.

\begin{flushleft}
  \textbf{Keywords:} general relativity, teleparallel gravity, tetrad formalism, action, regularisation, spin connection
\end{flushleft}

\newpage 
\pagenumbering{gobble}
\tableofcontents

\chapter*{Introduction}
 \markboth{INTRODUCTION}{INTRODUCTION}  
    \addcontentsline{toc}{chapter}{Introduction}
    \pagenumbering{arabic}

At the end of the 19th century, it was still unclear how to unify the new theory on electromagnetism - governed by the famous Maxwell's equations - with at that time still well-corroborated Newtonian mechanics. A few years later, on 26th of September 1905, Albert Einstein comes to the rescue with his paper "On the Electrodynamics of Moving Bodies"
\cite{Einstein:1905ve}, where he introduces his revolutionary special relativity. The key concept was to replace the good-old Galiean transformations, which implicitly assumed the universality of time, by the Lorentz transformations.

Thus, temporal and spatial dimensions were for the first time treated at an equal footing. Such rebellious, yet beautiful theory could be derived from two physically motivated postulates of principle of relativity and invariance of the speed of light, which lead to a plethora of interesting phenomena, such as length contraction, time dilation and mass-energy equivalence just to name a few. All of this ultimately allowed for compatibility of electromagnetism and classical mechanics.

In the upcoming years, Einstein began to incorporate gravity to this special relativistic picture \cite{Einstein:1916vd}. As it happens, the word "special" literally refers to "a special case" - that is, the theory is valid only in the special class of frames called inertial frames. Then, in the year 1915, Einstein presented his gravitational field equations, which directly relate the curvature of spacetime to the spatial configuration of matter, generalising thus the concept of relativity. As a result of this triumphant theory, gravity for the first time acquired purely geometrical interpretation in contrast to how we understand other fundamental forces of nature - in the sense of equivalence principle, objects that fall or orbit are not actually pulled by a force, but they rather follow straight-line-constant-speed paths - geodesics - in a curved spacetime. This is sometimes responsible for the widely accepted interpretation of the Einstein's field equations as: matter tells the spacetime how to curve, and in turn, curvature dictates how the matter moves \cite{Misner:1973prb}. It seemed that Einstein was correct yet again, not only explaining the so-far perplexing question of precession of Mercury, but presenting us with even more breath-taking predictions in the form of gravitational lensing, gravitational waves, and at last, but by no means the least, the eerie black holes.

Black hole; a place, where even light can not escape its gargantuan gravitational pull, remain in my opinion the least understood, and yet the most studied objects in physics. With direct experimental observations available only recently \cite{EventHorizonTelescope:2019dse}, these desolate entities are paradoxically used as an evidence for general relativity on one hand, and against its completeness on the other. The defining property is the so-called event horizon, the point of no return, with its origin in a singular point - the infinitely dense singularity resulting from gravitational collapse. Determining the nature of these singularities  is to this day one of the greatest issues in the classical general relativity \cite{Penrose:1969pc,Wald:1997wa,Penrose:1999vj}.

It turns out that any massive object, when compressed hard enough, becomes a black hole with all of its mass localised in the singularity. However, describing an infinitesimal object is within the bounds of quantum theory. This brings us to the greatest puzzle of modern theoretical physics, when the Cosmos at the largest scales is described by general relativity, while on the microscopic scales, quantum theory assumes command. To this day, no one has been successful in uniting these two theories and presenting us with a satisfactory model of quantum gravity, i.e. a theory of gravity that works even at the microscopic scales. Understanding the behaviour of black holes could thus provide us with hints at the reconciliation of these contrasting worldviews.

It turns out there were already several attempts at quantising gravity. The starting point is usually to specify the Lagrangian at the classical level of the particular model at hand, either from the prior knowledge of equations of motion, or based on the symmetries we expect from the theory. Then, one possible route is the canonical quantum gravity where we determine the Hamiltonian by the means of Legendre transformation. Such an approach, based on the Hamiltonian formalism, lead to the development of the loop quantum gravity \cite{Bojowald:2010qpa,Rovelli:1997yv}.

The other option is to remain within the Lagrange formalism and quantise the theory through the path integrals. This puts the emphasis on the action - an integral of the Lagrangian density. As the action of the Lorentzian signature is real and hence the path integral is oscillating, we typically employ the Wick's rotation, resulting in the Euclidean path integral formulation. Such a formulation was firstly introduced by none other than Hawking, who utilised it in his an alternative derivation of the black hole thermodynamics \cite{Hawking:1979ig,Hawking:1978jz}.

To this end, studying the gravitational black hole action seems to be of utmost interest. In the context of general relativity, as we shall see in the upcoming chapters, the standard Einstein-Hilbert action needs to be modified with a certain boundary term, coined as the Gibbons-Hawking-York (GHY) boundary term \cite{Gibbons:1976ue,York:1972sj}.  Nevertheless, naive addition of such a term causes the action to diverge. One thus needs to employ additional regularising methods when it comes to the actual evaluation of the action integral, with finite action needed for the definition of a variety of conserved charges, or thermodynamical quantities.

''But what does then the author mean by such a peculiar wording: Einstein's other gravity\footnote{This phrase was first adopted by Linder in \cite{Linder:2010py}}", one may ask. It turns out that Einstein himself did not stop at general relativity, but he wanted to further unify this new perspective on gravity with electrodynamics \cite{Einstein:1930xdd}; the notion, which started it all back in the late 1920s. Although he failed in this respect, his proposal of the idea of distant parallelism - or teleparallelism, was later revived in the 1960s as an alternative theory of gravity by Møller \cite{moller1961conservation}.

The central idea is, in simple terms, to reformulate general relativity in terms of torsion within teleparallel geometry instead of curvature in the Riemannian sense. From the mathematical viewpoint, this means replacing the Levi-Civita  connection by the so-called teleparallel connection. The other change in perspective is that with regards to teleparallel approach, it is natural to make use of the non-coordinate bases called as tetrads (vierbein - four-legged) in contrast to the coordinate ones.

This notion of teleparallelism gained in popularity in recent years, especially with respect to various modified theories of gravity aiming to address the perplexing nature of dark energy in the cosmological context \cite{Bahamonde:2021gfp,Krssak:2018ywd}. We shall, however, retain a bit more conservative viewpoint of teleparallel equivalent of general relativity (TEGR) \cite{Aldrovandi:2013wha}. In this particular formulation, teleparallel gravity is fully equivalent to general relativity with regards to the dynamical content of the theory. Nevertheless, the teleparallel Lagrangian does differ from the Lagrangian of general relativity by a total derivative term.

Thus, it will be precisely this one crucial distinction in the Lagrangian structure of the two gravitational frameworks at hand that we shall try to utilise in the subsequent regularisation of the black hole action. As we shall argue in the following chapters, teleparallel action does not need to be supplemented with a GHY-like boundary term. Instead, it requires us to determine a spin connection of a pure-gauge form. There is, however, a certain freedom in the actual specification of such a connection, as there are no additional field equations governing its own dynamics. Therefore, we shall try to make use of this freedom with regards to the action regularisation in order for us to obtain a valid prediction for the black hole action.

To this end, we shall investigate the procedure of the so-called ''background subtraction" used by Gibbons-Hawking (GH) and apply it to the gravitational action of the Schwarzschild black hole. In this way, however, we get information only about the exterior part of the black hole, as the singularity at the event horizon prohibits us to cover the whole of Schwarzschild spacetime. Nevertheless, there have been attempts to also evaluate the action inside the black hole in the context of holographic complexity by Susskind et al. \cite{Brown:2015bva,Brown:2015lvg}. Comparing this to the teleparallel approach, we will see that we obtain matching results in the interior region of the black hole. But when it comes to the external solution, teleparallel gravity interestingly predicts twice the result of the Gibbons-Hawking method. We shall thus analyse the root of this discrepancy by the employment of various frames: the so-called proper \cite{Krssak:2018ywd,Krssak:2023nrw} and canonical frames \cite{BeltranJimenez:2019bnx}. Additionally, we are going to explore the role of singular points in the action and check the validity of Stokes' theorem, which we use so abundantly in the practical setting.

\bigskip

{\it{Notation:}} Due to us considering not one, but as much as three distinct connections, we need to somehow differentiate between the geometrical objects related to each of them to make everything as comprehensible and cohesive as possible. To this end, we shall denote quantities related to the teleparallel connection with a bare text, with no stylistic modifications whatsoever, due to the sheer frequency such objects will be used with. On the other hand, quantities with regards to a general metric-affine connection will be represented with a bold text. Lastly, objects with respect to the Riemannian connection will be adorned with a $"\circ"$ symbol on top of them. In this sense, $\Gamma^\rho{}_{\mu \nu}, \Gammag^\rho{}_{\mu \nu}, \Gammac^\rho{}_{\mu \nu}$ represent the linear connection of teleparallel, general metric-affine, and Riemannian geometries respectively. The Greek indices will be reserved for the components of tensors with respect to the coordinate bases, while Latin indices will denote the components in the non-coordinate bases. The Lorentzian signature is assumed to be $(+,-,-,-)$ throughout the whole thesis. 

\chapter{Geometrical background}
Here we present the geometrical framework for the topic of interest. Setting the stage with the standard tetrad formalism, we then move through the definitions of the general affine connection and the spin connections in the general context of metric-affine geometries, to end up with the two restricted cases of the Riemannian and teleparallel connection. Lastly, we present a geometrical interpretation of the two main tensors characterising the connection: curvature and torsion tensors. 

\section{Tetrad formalism \label{chaptetradformalism}}
As has already been hinted at in the Introduction, the standard formalism adopted when dealing with the teleparallel framework is the language of tetrads. As we shall see shortly, tetrads represent a straightforward generalisation of coordinate bases to non-coordinate ones. In this respect, we shall mainly draw upon the following works \cite{Aldrovandi:2013wha,Krssak:2018ywd,Carroll:2004st,Krssak:2024xeh}.

Similarly as in general relativity, for the underlying base manifold we take the 4-dimensional spacetime $\mathcal{M}$ with the Minkowski spacetime $\mathbb{R} ^{1,3}$ comprising the tangent space at each point. For the spacetime indices we adopt the use of Greek letters ($\mu,\nu...=0,1,2,3$), and the tangent space indices shall be represented by Latin alphabet ($a,b...=0,1,2,3$), with the Minkowski metric in Cartesian coordinates assuming the form $\eta_{a b} = \text{diag}(+1,-1,-1,-1)$. At each point p of the manifold $\mathcal{M}$, we can choose a set of local coordinates ${x^\mu}$, which define a coordinate basis $\{ \partial_\mu\}$ and $\{ dx^\mu\}$ for the vector or covector fields respectively as
\begin{equation}
    V=V^\mu \partial_\mu, \quad \alpha=\alpha_\mu dx^\mu,
\end{equation}
where $V \in \mathcal{T}^1_0(\mathcal{M})$ stands for a general vector field and $\alpha \in \mathcal{T}^0_1(\mathcal{M})$ represents a general covector field. A general tensor field $X$ of type $(^p_q)$ ($X\in \mathcal{T}^p_q(\mathcal{M})$) can thus be decomposed with respect to the coordinate basis as
\begin{equation}
    X=X^{\mu...\nu}{}_{\rho ...\sigma} \partial_\mu \otimes...\otimes \partial_\mu \otimes dx^\rho \otimes ... \otimes dx^\sigma.
\end{equation}
The defining property of tensor fields is that they are invariant under a change of coordinates
\begin{equation}
    x^\mu \mapsto x'^\mu(x),
\end{equation}
which translates to the following transformation rules for the vector basis
\begin{equation}
    \partial_\mu \mapsto \partial_{\mu}{}'\equiv\frac{\partial x^\nu}{\partial x'^\mu}\partial_\nu,
\end{equation}
and covector basis
\begin{equation}
    dx^\mu \mapsto dx'^\mu\equiv\frac{\partial x'^\mu}{\partial x^\nu}dx^\nu.
\end{equation}
This then allows us to determine the transformation property of the tensor components
\begin{equation}
    X'^{\mu...}{}_{\nu...}=X^{\rho...}{}_{\sigma ...} \frac{\partial x'^\mu}{\partial x^\rho}...\frac{\partial x^\sigma}{\partial x'^\nu}...
\end{equation}
Nevertheless, we may consider a more general class of bases by taking a set of four linearly independent vectors $e_a$ to define the so-called non-coordinate basis (frame field) for the vector fields in the tangent space. Similarly, we may define a non-coordinate covector basis $e^a$ in the cotangent space satisfying the orthogonality condition
\begin{equation}
    e^a{}_\mu e_b{}^\mu = \delta^a_b.
\end{equation}
These basis vectors and covectors can be decomposed with respect to the coordinate basis straightforwardly as
\begin{equation}
    e_a={e_a}^\mu \partial_\mu, \quad e^a={e^a}_\mu dx^\mu.
\end{equation}
Moreover, for a general frame field $e_a$, the following commutation relation holds
\begin{equation}
    [e_a, e_b]={f^c}_{a b} e_c,
\end{equation}
with ${f^c}_{a b}$ being the coefficients of anholonomy, with the dual relation given as
\begin{equation}\label{cartstreq}
    de^c=-\frac{1}{2} {f^c}_{a b}e^a \wedge e^b=\frac{1}{2}(\partial_\mu {e^c}_\nu-\partial_\nu {e^c}_\mu)dx^\mu \wedge dx^\nu.
\end{equation}
This then allows for an expression for the coefficients of anholonomy in terms of a given frame field ${e^a}_\mu$ as 
\begin{equation}
    {f^c}_{a b}={e_a}^\mu {e_b}^\nu (\partial_\nu {e^c}_\mu - \partial_\mu {e^c}_\nu).
\end{equation}
In the case that the basis vectors do not commute, i.e. $f^c{}_{a b} \neq 0$, it is not possible to express the frame $e^a$ in terms of some coordinates, hence the name non-coordinate basis. However, it is possible that locally, these coefficients vanish identically ${f^c}_{a b}=0$, which causes the eq. \eqref{cartstreq} to resolve trivially.  Then, according to Poincaré lemma\footnote{It holds that for any contractible manifold (and therefore locally for any manifold), a closed differential form is also an exact form $d\alpha=0 \Rightarrow \alpha = d\beta$ \cite{Fecko:2006zy}.}, we speak about the so-called holonomic bases; at an infinitesimal neighbourhood of a point $p$ there always exist local coordinates ${x^a}$ such that
\begin{equation}\label{inertialframe}
    e^a \equiv dx^a, \quad  e_a = \frac{\partial}{\partial x^a}\equiv \partial_a.
\end{equation}
In the case of special relativity, when the spacetime metric $g_{\mu \nu}$ assumes the form of the Minkowski metric $\eta_{\mu \nu}$, frames of this particular holonomic form are named as inertial frames \cite{Aldrovandi:2013wha}.

To express the components of an arbitrary tensor in the non-coordinate basis, all we need is to consider the following contractions
\begin{equation}
    X^{a...}{}_{b...}=X^{\mu...}{}_{\nu...}e^a{}_\mu ... e_b{}^\nu...
\end{equation}
An important observation here is that tensor components with respect to the non-coordinate basis are invariant under coordinate transformations, as $e^a{}_\mu$ is a coordinate covector in the last index. On the other hand, we may consider a general change of basis 
\begin{equation}\label{genframetsf}
    e^a \mapsto e'^a=B^a{}_b e^b,
\end{equation}
with $B^a{}_b\in GL(4,\mathbb{R})$ being a general invertible matrix. This time around, the tensor components in the non-coordinate basis do change as
\begin{equation}
    X'^{a...}{}_{b...}=X^{c...}{}_{d...} B^a{}_c...(B^{-1})^d{}_b
\end{equation}

Next we introduce the metric tensor $g=g_{\mu \nu}dx^\mu \otimes dx^\nu$ on a manifold, which allows for the definition of a scalar product of two vector fields and hence to measure distances and angles, resulting in a (pseudo)-Riemannian manifold $(\mathcal{M}, g)$. The metric components can be expressed in the non-coordinate basis as follows
\begin{equation}
    g_{a b}=g_{\mu \nu} e_a{}^\mu e_b{}^\nu.
\end{equation}
A particularly interesting class of non-coordinate basis turns out to be that of orthonormal basis vectors. This case is achieved by fixing the metric in the tangent space to be the Minkowski metric $\eta_{a b}$. Then we speak about the tetrad field, or tetrads in short, denoted by $h^a$ throughout the whole thesis. The relation between metric components in the coordinate basis and the non-coordinate one thus becomes
\begin{equation}
    g_{\mu \nu}=\eta_{a b} h^a{}_\mu h^b{}_\nu.
\end{equation}
At the same time, the class of transformations of an orthonormal frame $h^a$
\begin{equation}
    h^a \mapsto h'^a=\Lambda^a{}_b h^b,
\end{equation}
is restricted to such transformations that leave the tangent space Minkowski metric $\eta_{a b}$ invariant
\begin{equation}\label{tetradtsf}
    \eta_{a b}=\Lambda^c{}_a(x)\Lambda^d{}_b(x)\eta_{cd},
\end{equation}
i.e. the local Lorentz transformations $\Lambda^a{}_b(x) \in SO(1,3)$.

\section{General affine connection}
In this section, we shall introduce the concept of a general affine connection $\nablag$ and the notion of the covariant derivative, which essentially generalises the concept of derivatives of tensors on "curved" manifolds. Throughout this section, $\nablag$ is to be understood in the sense of metric-affine geometry, with the specific limiting cases considered afterwards. We shall draw mainly upon the following standard textbooks on differential geometry \cite{Fecko:2006zy,Nakahara:2003nw}.

For our abstract playground, we take the differentiable manifold $\mathcal{M}$ endowed with the structure of an affine connection $\nablag$. This additional structure allows for a definition of parallel transport of a general tensor field $X \in \mathcal{T}_{q}^{p}(M)$ along an arbitrary curve  $\gamma$ by the following coordinate-free prescription 
\begin{equation}
\nablag_{\dot{\gamma}} X := 0,
\end{equation}
which is really a condition for the tensor field X to be autoparallel; that is, we take the tensor at the initial position $X^{\mu...}{}_{\nu...}(t_1)$, create its "parallel" copies along the whole path via solving the differential equations of parallel transport, and finally the transported tensor is found to be $X^{\mu...}{}_{\nu...}(t_2)$. Notice that in this way, the covariant derivative of an object informs us about the deviation of the field from being parallel.

In order for us to derive the actual equations of parallel transport, we need to specify the rule for the covariant derivative of a general frame field $e_a$ (i.e. not necessarily orthonormal frames), which is given uniquely by a collection of functions known as the coefficients of linear connection $\Gammag^a{}_{b c}(x)$
\begin{equation}
    \nablag_{a}e_b =: \Gammag^c{}_{b a}e_c, \quad\text{where}\quad \nablag_{a} \equiv \nablag_{e_a}.
\end{equation}
Utilising the $\mathcal{F}$-linearity \footnote{$\mathcal{F}(M)$-linearity, in loose terms, generalises the $\mathbb{R}$-linearity. That is, the mapping is linear not only with respect to a real number, but rather to any function $f \in \mathcal{F}(M)$.}
\begin{equation}
    \nablag_{V+fW}=\nablag_V+f\nablag_W,
\end{equation}
we may now write the covariant derivative of a frame $e_a$ along arbitrary vector field $W=W^a e_a$ as
\begin{equation}\label{christalg}
    \nablag_{W}e_b \equiv \nablag_{(W^a e_a)} e_b =W^a \nablag_{a}e_b= \Gammag^c{}_{b a} W^a e_c.
\end{equation}
For holonomic (coordinate) frames $e_\mu \equiv \partial_\mu$, we adopt an analogous prescription
\begin{equation}\label{christdef}
    \nablag_{\nu} \partial_\mu =: \Gammag^\rho{}_{\mu \nu} \partial_\rho, \quad\text{where}\quad \nablag_{\nu} \equiv \nablag_{\partial_\nu}.
\end{equation}
Using now the properties that the covariant derivative acts on functions $f \in \mathcal{F}(M)$ as an ordinary derivative
\begin{equation}
    \nablag_W f = W f,
\end{equation}
and it commutes with contractions 
\begin{equation}
    0=\nablag_{W}\delta^a_b=\nablag_{W}(C(e^a \otimes e_b))\equiv C(\nablag_{W}(e^a \otimes e_b)),
\end{equation}
we find that the covariant derivative of a co-frame field $e^a$ is realised as
\begin{equation}
    \nablag_{a}e^b = -\Gammag^b{}_{c a}e^c, 
\end{equation}
or in local coordinates
\begin{equation}
    \nabla_{\nu}dx^\rho = -\Gammag^\rho_{\mu \nu}dx^\mu.
\end{equation}
Assuming that the curve $\gamma$ happens to be an integral curve of vector field W ($W=\dot{\gamma}$) and restricting ourselves for simplicity to the case of a vector field $V \in \mathcal{T}^1_0(M)$, we arrive at the expression for equations of parallel transport in local coordinates
\begin{equation}
    \nablag_{W}V = ( \dot{V^\rho}+\Gammag^\rho{}_{\mu \nu} \dot{x^\mu} V^\nu) \partial_\rho =0.
\end{equation}
When we parallel transport objects, we may be inclined to postulate some further requirements, such as for the length or the relative angles between vectors to be preserved. Then we talk about the compatibility of structures; in this case the compatibility of linear connection $\nablag$ with the metric $g$. In coordinate-free language, the metric-compatibility condition is as follows
\begin{equation}
    \nablag g = 0,
\end{equation}
or in local coordinates
\begin{equation}\label{metriccomp}
    \Qg_{\rho \mu \nu} \equiv g_{\mu \nu;\rho} := \nablag g(\partial_\mu,\partial_\nu,\partial_\rho) \equiv \nablag_\rho g(\partial_\mu, \partial_\nu)=0,
\end{equation}
where we have defined the covariant gradient as a covariant derivative in an unspecified direction
\begin{equation}
    (\nablag_W A)^{\mu...\nu}{}_{\rho...\sigma}=W^\gamma A^{\mu...\nu}{}_{\rho...\sigma;\gamma},
\end{equation}
and the so-called non-metricity tensor $ \Qg_{\rho \mu \nu}$.
The interpretation of eq. \eqref{metriccomp} is to take the covariant derivative of a scalar product of any two vectors along an arbitrary path and let it vanish. Such a connection is then awarded a title: a metric compatible connection. By defining 
\begin{equation}
    \Gammag_{\rho \mu \nu}:=g_{\rho \sigma}\Gammag^\sigma{}_{\mu \nu},
\end{equation}
we find the metric compatibility condition to be equivalent to
\begin{equation}
    g_{\mu \nu, \rho}= \Gammag_{\mu \nu \rho}+\Gammag_{\nu \mu \rho}.
\end{equation}

Next, let us stop by the concept of geodesics. Realising that the acceleration is commonly defined as the rate of change of velocity, it should not come as a surprise that we would generalise it as the covariant derivative of velocity along the velocity itself
\begin{equation}
    a=\nablag_{\dot{\gamma}} \dot{\gamma}, \equiv \nablag_v v, \quad \text{where} \quad v := \dot{\gamma}.
\end{equation}
In other words, we are asking about the result of a parallel transport of a velocity vector $v$ along the path defined by the velocity vector. Setting this relation equal to zero thus results in the uniform straight-line motion generalised to curved spaces in an analogy to $\frac{dv}{dt}=0$ in classical mechanics. In local coordinates, this is equivalent to the following set of equations
 \begin{equation}
     \nablag_{\dot{\gamma}} \dot{\gamma} = 0 \quad \Leftrightarrow \quad \ddot{x}^\rho + \Gammag^\rho_{\mu \nu}\dot{x}^\mu \dot{x}^\nu =0.
 \end{equation}

Now let us inspect how the coefficients of linear connection in non-coordinate basis $\Gammag^a{}_{b c}$ respond to the change of frame \eqref{genframetsf} with a general non-singular matrix $B^a{}_b$
\begin{equation}\label{lincontsfa}
    \Gammag'^c{}_{ab}=\Gammag^d{}_{ef}(B^{-1})^c{}_d B^e{}_a B^f{}_b +(B^{-1})^c{}_d B^f{}_b (e_f B^d{}_a).
\end{equation}
Or equivalently, the transformation rule for the connection coefficients in terms of the coordinate basis $\Gammag^\rho_{\mu \nu}$
\begin{equation}\label{lincontsfb}
    \Gammag'^\mu{}_{\nu \rho}=\Gammag^\alpha{}_{\beta \sigma} \frac{\partial x'^\mu}{\partial x^\alpha} \frac{\partial x^\beta}{\partial x'^\nu} \frac{\partial x^\sigma}{\partial x'^\rho} + \frac{\partial x'^\mu}{\partial x^\alpha}\frac{\partial^2 x^\alpha}{\partial x'^\nu \partial x'^\rho}.
\end{equation}
In both equations \eqref{lincontsfa} and \eqref{lincontsfb} we see a "good" term - one that transforms covariantly as a $(^1_2)$ tensor - but there is also a second, inhomogeneous term. The coefficients of linear connection with respect to both coordinate and non-coordinate bases thus fail to be tensorial in nature. Nevertheless, it is precisely this inhomogeneity in transformation of connection coefficients that ensures that the covariant derivative of tensor fields as a resulting object transforms covariantly.

\section{Spin connection}

In differential geometry, we may sometimes want to switch to the language of differential forms. Namely, instead of working with the connection coefficients $\Gammag^\rho{}_{\mu \nu}$, we may introduce their counterpart; the connection 1-forms. In the case of a general metric-affine connection, we thus define the matrix valued 1-form \cite{Fecko:2006zy} with respect to a general co-frame field $e^a={e^a}_\mu dx^\mu$ as
\begin{equation}
    \omegag^a{}_b \equiv \omegag^a{}_{b\mu} dx^\mu,
\end{equation}
based on the following identification
\begin{equation}
    \nablag_V e_b = \omegag^a{}_b (V) e_a \iff \omegag^a{}_b = \Gammag^a{}_{b c} e^c. 
\end{equation}

In the special case, when the frame field is orthonormal, i.e. instead of a general frame we work with tetrads $e^a \rightarrow h^a$, and the connection $\nablag$ is metric compatible, we tend to call the connection 1-form $\omegag^a{}_{b \mu}$ as a spin connection \cite{Aldrovandi:2013wha,Krssak:2024xeh}. While in the coordinate basis, it is natural to work with a pair $\{g_{\mu \nu}, \Gammag^\rho{}_{\mu \nu} \}$ - in the case of orthonormal non-coordinate basis, it is practical to make use of a pair $\{ h^a{}_\mu, \omegag^a{}_{b \mu} \}$.

In this way, we may talk about covariant derivative $\nablag$ acting on components of vectors expressed in two separate bases
\begin{equation}
    X^a{}_{;\mu} = \partial_\mu X^a+\omegag^a{}_{b \mu}X^b, \quad  X^\rho{}_{;\mu} = \partial_\mu X^\rho+\Gammag^\rho{}_{\nu \mu}X^\nu,
\end{equation}
with the spin connection coefficient $\omegag^a{}_{b \mu}$ acting on components in non-coordinate basis, while the linear connection coefficients $\Gammag^\rho{}_{\mu \nu}$ acts on tensors in coordinate basis.

The key point is that these are simply two different connection coefficients for the same connection $\nablag$ acting on components of tensors expressed in two different bases. This consequently allows for a relation between the two kinds of connection coefficients, which stems from the following decomposition of a general vector $X$ being expressed in coordinate and non-coordinate basis as 
\begin{equation}
    X=X^\mu \partial_\mu = X^a h_a.
\end{equation}
Realising now that the covariant derivative of this vector $X$ must be the same, regardless of the basis, it is expressed as
\begin{equation}
    \nablag_\mu X = ( \nablag_\mu X^\nu ) \partial_\nu = (\nablag_\mu X^a) h_a,
\end{equation}
we get the desired relation between the spin connection $\omegag^a{}_{b \mu}$ in terms of the spacetime indexed connection coefficients $\Gammag^\rho{}_{\mu \nu}$ as
\begin{equation}
    {\omegag^a{}_b}_\mu = {h^a}_\rho \Gammag^\rho{}_{\mu \nu} {h_b}^\nu + {h^a}_\rho \partial_\mu {h_b}^\rho.
\end{equation}
Or equivalently, we get the expression for the coefficients of the affine connection
\begin{equation}
    {\Gammag^\rho}_{\nu \mu}= {h_a}^\rho  {\omegag^a{}_b}_\mu {h^b}_\nu + {h_a}^\rho \partial_\mu {h^a}_\nu.
\end{equation}
To investigate the transformation properties of the spin connection, all we need is to realise that in contrast to the change of a general frame in eq. \eqref{genframetsf}, the class of transformations of tetrad frame $h^a$ are the local Lorentz transformations $\Lambda^a{}_b(x)$. The spin connection consequently responds to a change of a tetrad \eqref{tetradtsf} as
\begin{equation}
   {\omegag^a{}_b}_\mu \rightarrow {{\omegag'}^a{}_b}_\mu=\Lambda^a{}_c {\omegag^c{}_d}_\mu \Lambda_b{}^d + \Lambda^a{}_c\partial_\mu \Lambda_b{}^c,
\end{equation}
where we have denoted the inverse Lorentz transformation as $(\Lambda^{-1})^a{}_b \equiv \Lambda_b{}^a$. This transformation property of the spin connection thus ensures that the new objects, when expressed in a orthonormal non-coordinate basis, transform covariantly under the local Lorentz transformations.

Moreover, if we recall the metricity condition \eqref{metriccomp}, now understood in the sense of tangent space Minkowski metric
\begin{equation}
    \eta_{ab; \mu}=\eta_{ab,\mu}-{\omega^c{}_a}_\mu \eta_{c b}-{\omega^c{}_b}_\mu \eta_{a c} = 0,
\end{equation}
it follows that
\begin{equation}
    \omega_{a b \mu}=-\omega_{b a \mu}.
\end{equation}
That is, it leads to the spin connection being antisymmetric in the tangent space indices. In geometers jargon, we say that the spin connection assumes values in the Lie algebra of the Lorentz group. From now on we may thus use the names algebraic and tangent space indices interchangeably.

In the most general case of metric-affine geometries, both connection coefficients have $4^3 = 64$ independent components, and thus they can be characterised in terms of the three following tensors: curvature, torsion, and non-metricity. The corresponding definitions with respect to coordinate basis are
\begin{equation}\label{riemanncurvature}
    \Rg^\rho{}_{\mu \nu \sigma}:=\Gammag^\rho{}_{\mu \sigma,\nu}-\Gammag^\rho{}_{\mu \nu,\sigma}+\Gammag^\gamma{}_{\mu \sigma}\Gammag^\rho{}_{\gamma \nu}-\Gammag^\gamma{}_{\mu \nu}\Gammag^\rho{}_{\gamma \sigma},
\end{equation}
\begin{equation}
    \Tg^\rho{}_{\mu \nu} := \Gammag^\rho{}_{\nu \mu}-\Gammag^\rho{}_{\mu \nu} ,
\end{equation}
\begin{equation}
    \Qg_{\rho \mu \nu}:= \nablag_\rho g_{\mu \nu}.
\end{equation}

These expressions can be also rewritten with regards to the non-coordinate basis as
\begin{equation}\label{curvatureform}
     \Rg^a{}_{b \mu \nu}:= \omegag^a{}_{b \nu, \mu}-\omegag^a{}_{b \mu, \nu} +\omegag^a{}_{c \mu}\omegag^c{}_{b \nu}-\omegag^a{}_{c \nu}\omegag^c{}_{b \mu},
\end{equation}
\begin{equation}\label{torsion}
    {\Tg^a}_{\mu \nu} := \partial_\mu {h^a}_\nu - \partial_\nu {h^a}_\mu + {\omegag^a{}_b}_\mu {h^b}_\nu - {\omegag^a{}_b}_\nu {h^b}_\mu,
\end{equation}
\begin{equation}
    \Qg_{\mu a b}:= \nablag_\mu \eta_{a b} = \omegag_{a b \mu} + \omegag_{b a \mu}.
\end{equation}

\section{Riemannian connection}
Riemannian or Levi-Civita (RLC) connection is the most widely used connection, defined by the metric compatibility and symmetricity condition (vanishing torsion). In the coordinate basis, these two conditions give rise to the connection coefficients given uniquely in terms of the metric as
\begin{equation}\label{RLCchrist}
    \Gammac^\rho{}_{\mu \nu}=\frac{1}{2} g^{\rho \sigma} (g_{\sigma \mu,\nu}+g_{\sigma \nu, \mu}-g_{\mu \nu, \sigma}),
\end{equation}
usually known under the name of Christoffel symbols of the second kind. Similarly, the RLC spin connection of the non-coordinate basis is given uniquely, this time in terms of the coefficients of anholonomy by
\begin{equation}
    \omegac^a{}_{b \mu}=\frac{1}{2} h^c{}_\mu \left( f_b{}^a{}_c+f_c{}^a{}_b-f^a{}_{b c} \right).
\end{equation}

\section{Teleparallel connection}
Next, we present the form of the teleparallel spin connection used throughout the rest of the thesis. Such a connection is defined by demanding the metric compatibility, and vanishing curvature. The latter condition can be shown to be fulfilled by a pure gauge connection
\begin{equation}
    \omega^a{}_{b \mu}=\Lambda^a{}_c \partial_\mu \Lambda_b{}^c,
\end{equation}
with $\Lambda^a{}_b(x)$ belonging to the class of local Lorentz transformations $SO(1,3)$ after insisting on the metric compatibility.

The spacetime-indexed coefficients of teleparallel connection are then related as
\begin{equation}\label{conneccoeftg}
    {\Gamma^\rho}_{\nu \mu}= {h_a}^\rho  {\omega^a{}_b}_\mu {h^b}_\nu + {h_a}^\rho \partial_\mu {h^a}_\nu.
\end{equation}

Particularly useful is the so-called Ricci theorem, relating the teleparallel spin connection to the RLC spin connection
\begin{equation}\label{telspincondec}
    {\omega^a}_{b \mu}=\omegac^a{}_{b \mu}+{K^a}_{b \mu},
\end{equation}
and the contortion tensor 
\begin{equation}
     K^a{}_{b \mu}=\frac{1}{2}({{T_\mu}^a}_b+{{T_b}^a}_\mu-{T^a}_{b \mu}).
\end{equation}

On the other hand, the spacetime-indexed teleparallel connection coefficients decompose into already familiar Christoffel symbols
\begin{equation}
    \Gammac^\rho{}_{\mu \nu}=\frac{1}{2} g^{\rho \sigma} (g_{\sigma \mu, \nu}+g_{\sigma \nu, \mu}-g_{\mu \nu, \sigma}),
\end{equation}
and the contortion tensor
\begin{equation}
    {K^\rho}_{\mu \nu}=\frac{1}{2}({{T_\mu}^\rho}_\nu+{{T_\nu}^\rho}_\mu-{T^\rho}_{\mu \nu}),
\end{equation}
as
\begin{equation}
    \Gamma^\rho{}_{\mu \nu}=\Gammac^\rho{}_{\mu \nu} + {K^\rho}_{\mu \nu}.
\end{equation}

\section{Geometrical interpretation}
Lastly, here we try to motivate the two main tensors, curvature and torsion, in a geometrical way \cite{Fecko:2006zy}. Starting off with the coordinate-free definition of curvature tensor
\begin{equation}
    \Rg(W,U,V;\alpha) := \langle \alpha, R(U,V)W \rangle \equiv \langle \alpha, ([\nablag_U,\nablag_V]-\nablag_{[U,V]})W \rangle,
\end{equation}
where $U,V,W \in \mathcal{T}^1_0(\mathcal{M})$ are arbitrary vector fields, $\alpha \in \mathcal{T}^0_1(\mathcal{M})$ is a general 1-from, and $\Rg(U,V)$ is the curvature operator defined as
\begin{equation}
    \Rg(U,V):=[\nablag_U,\nablag_V]-\nablag_{[U,V]}.
\end{equation}
One might arrive at such a definition along subsequent lines: we take two vector fields U,V. Then we take an arbitrary object situated at a point A and parallel transport it along U by an infinitesimal parameter $\epsilon$ to a point B, with the operator of parallel transport from A to B given as 
\begin{equation}
    \tau_{B,A}=e^{-\epsilon\nablag_U}=\hat{1}-\epsilon\nablag_U+\frac{1}{2}\epsilon^2 \nablag^2_U+...
\end{equation}
 Now we transport the same object from the point B along vector field V to a point C; from where we try to close the loop by performing the same sequence of steps in the negative directions of given vector fields. In this way, we find that the expression for the parallel transport along such a loop is with the second-order accuracy in $\epsilon$ given as
\begin{equation}   
    \tau_{A,A}=e^{\epsilon^2\nablag_{[U,V]}}e^{\epsilon\nablag_V}e^{\epsilon\nablag_U}e^{-\epsilon\nablag_V}e^{-\epsilon\nablag_U}=\hat{1}-\epsilon^2\Rg(U,V)+...
\end{equation}

\begin{figure}[!ht]
\centering
\includegraphics[width=.5\linewidth]{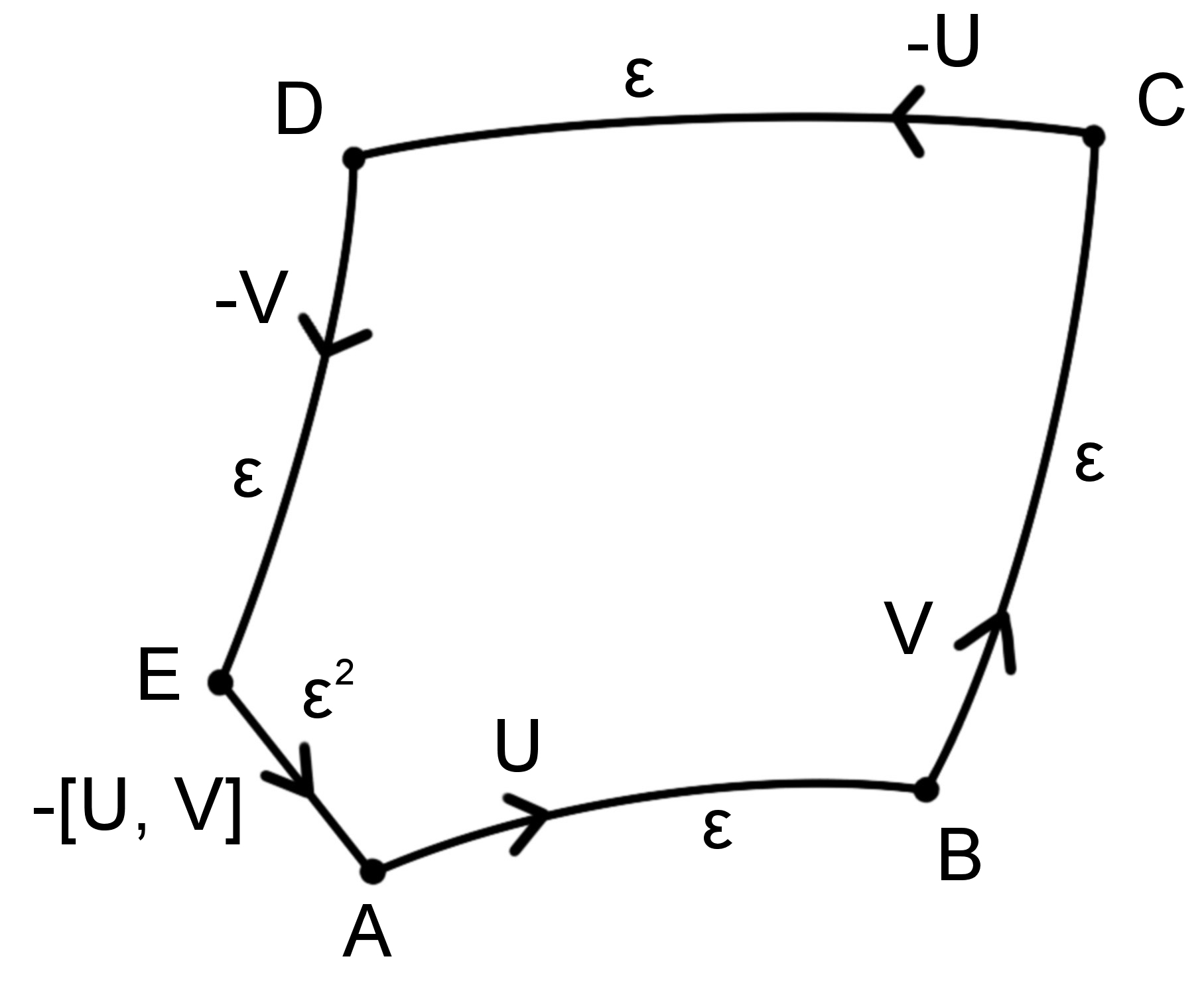}
\captionof{figure}{Paths taken under the parallel transport.}
\end{figure}

We thus see that due to the path dependence of parallel transport, the transported object along an infinitesimal closed loop suffers a change proportional to the curvature operator, i.e. if we were to assume Riemannian connection, the transported tensor undergoes a rotation.

The coordinate-free definition of torsion tensor is as follows
\begin{equation}
    \Tg(V,W) := \nablag_V W - \nablag_W V - [V,W],
\end{equation}
where $V,W \in \mathcal{T}^1_0(\mathcal{M})$ are vector fields. To illustrate the effects of torsion in a geometrical way, we may consider a similar situation as before. The only difference now is that instead of thinking of a change of an object at the level of tensors being transported, torsion emerges at the level of points along which we travel. Indeed, let us consider two vector fields $u,v$ situated at a point A. The point A and vector $u$ then define a unique (affinely parametrised) geodesic $\gamma_u (t)$. Now, we take the other vector $v$ and parallel transport it to the point $B\equiv\gamma_u (\epsilon)$ along this geodesic, which leads to a parallel vector $v_{||}$ in B. This in turn defines yet another geodesic $\gamma_{v_{||}}(t)$, with the point C defined as $C\equiv\gamma_{v_{||}}(\epsilon)$. Now we only need to repeat the whole procedure with the roles of vectors $u,v$ interchanged, obtaining the points D and E in the process. It is save to say that in the case of an ordinary plane, we ought to draw a parallelogram with the vertices given as A,B,$C\equiv E$,D. However, on a general manifold endowed with a connection $(\mathcal{M},\nablag)$, there holds $C\neq E$. Thus, to enclose the parallelogram, one ought to consider the last step given by the vector $T(u,v)$ within the second-order accuracy in $\epsilon$. The situation is as follows

\begin{figure}[!ht]
\centering
\includegraphics[width=.5\linewidth]{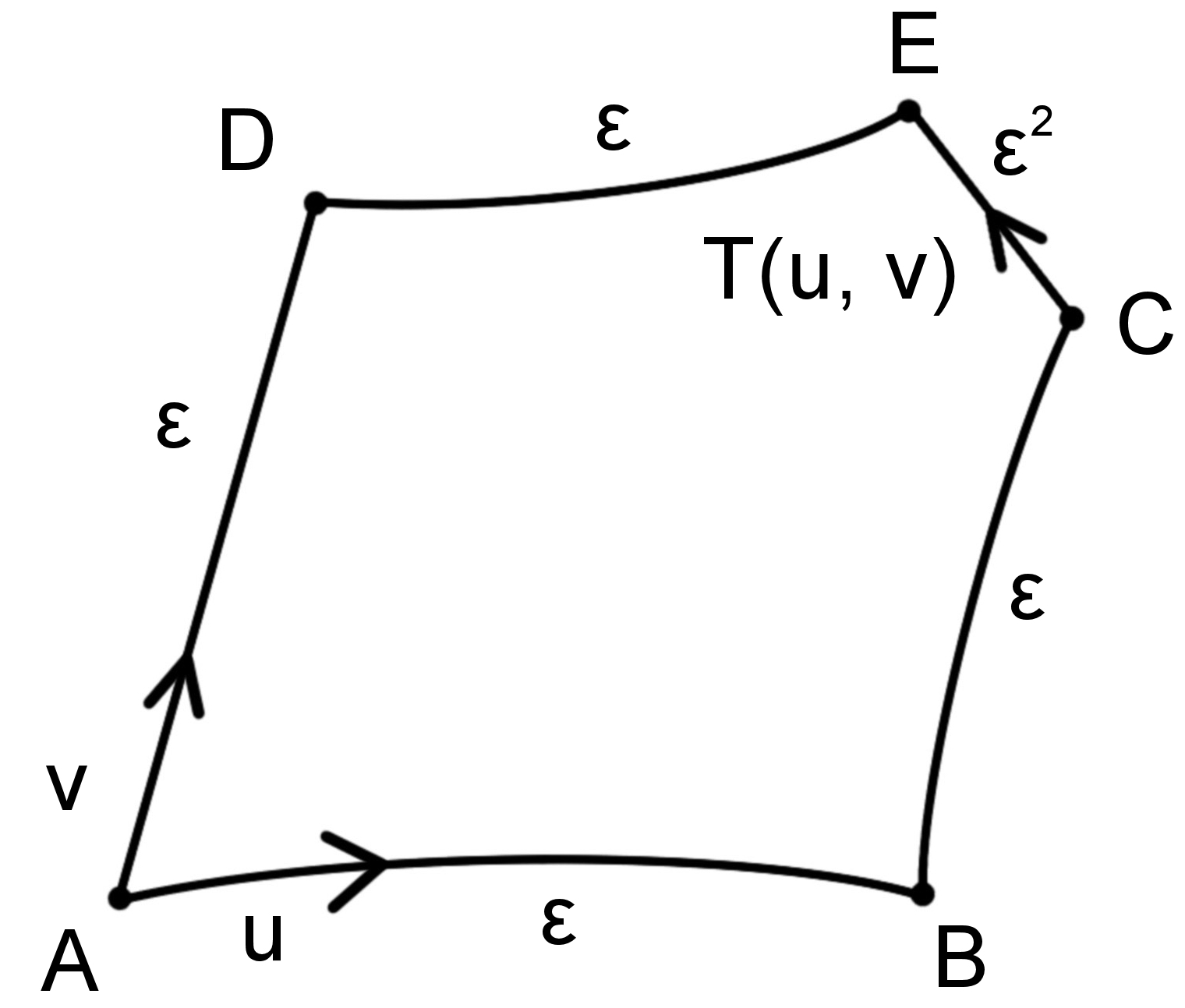}
\captionof{figure}{Vectors $u$ and $v$ with respective geodesics.}
\end{figure}

We thus see that both curvature and torsion tensors carry an important piece of information characterising the affine connection $\nablag$ on a general manifold $\mathcal{M}$. In the case of curvature, the transported tensors suffer a change due to the local path dependence of parallel transport. On the other hand, torsion causes a disclosure of a geodesic parallelogram. 

\chapter{Gravitational action}

In the upcoming sections, we shall present the central object of interest within the respective theories of general relativity and its teleparallel equivalent: the gravitational action. Adopting the Lagrange formalism, we shall derive the equations of motion governing the dynamics of both theories. We then discuss action and its regularisation in the Minkowski case. This prepares us for the debate regarding the regularisation in the case of Schwarzschild black hole in chapter~\ref{bhaction}.

\section{Gravitational action of general relativity}

Let us consider the following expression for the total gravitational action of general relativity \cite{Poisson:2009pwt}
\begin{equation}\label{naiveghy}
    \Ssc_{\text{tot}}=\Ssc_{\text{EH}}+\Ssc_{\text{GHY}}=-\frac{1}{2\kappa}\int_{\mathcal{M}}^{} d^4x \sqrt{-g} \Rc - \frac{1}{\kappa}\int_{\partial \mathcal{M}}^{} d^3y\, \epsilon \sqrt{|\gamma|} \Kc. 
\end{equation}
The first term above is the standard Einstein-Hilbert action \cite{Misner:1973prb}
\begin{equation}
    \Ssc_{\text{EH}}= \frac{1}{2\kappa}\int_{\mathcal{M}}^{} d^4x  \sqrt{-g} \Rc, 
\end{equation}
with $\Rc$ denoting the so-called Ricci scalar, the contraction of Ricci tensor $\Rc_{\mu \nu}$ with the spacetime metric $g_{\mu \nu}$
\begin{equation}\label{riccisc}
    g^{\mu \nu} \Rc_{\mu \nu}= \Rc^\mu{}_\mu \equiv \Rc,
\end{equation}
with the Ricci tensor itself being the contraction of Riemann curvature tensor $\Rc^\alpha{}_{\mu \beta \nu}$ in a following sense 
\begin{equation}
    \delta^\beta_{\alpha } \Rc^\alpha{}_{\mu \beta \nu} = \Rc^\beta{}_{\mu \beta \nu} \equiv \Rc_{\mu \nu}.
\end{equation}
Lastly, $g$ denotes the determinant of the metric $g\equiv \text{det} (g_{\mu \nu})$, and $\kappa \equiv 8 \pi$, with the gravitational constant and speed of light set as $G=c=1$ in the sense of natural units throughout the whole thesis.

The second term is known as the Gibbons-Hawking-York boundary term \cite{Gibbons:1976ue,York:1972sj}
\begin{equation}
    \Ssc_{\text{GHY}} = \frac{1}{\kappa}\int_{\partial \mathcal{M}}^{} d^3y\, \epsilon \sqrt{|\gamma|} \Kc.
\end{equation}
 The name ''boundary" stems from a simple fact that it in contrast to the Einstein-Hilbert term, which is essentially a volume integral over some spacetime region $\mathcal{M}$, the GHY term is evaluated across its $(D-1)-$dimensional boundary $\partial \mathcal{M}$. $\gamma_{i j}$ is then the metric induced on this boundary, with $\gamma$ being its determinant and $y^i$ are the coordinates on the boundary. $\Kc \equiv \gamma^{\mu \nu} n_{\mu ;\nu}= n^\mu{}_{;\mu}$ stands for the trace of the extrinsic curvature with $n^\mu$ being a vector normal to the boundary, parameter $\epsilon$ is either equal to $+1$ if the normal to $\partial \mathcal{M}$ is timelike and $-1$ when the normal to $\partial \mathcal{M}$ is spacelike.

The need for the consideration of the GHY boundary term thus arises whenever the underlying manifold $\mathcal{M}$ has a boundary, and  is required by the well-posedness of the variational principle \cite{Poisson:2009pwt}. This can be easily understood from the variation of the standard Einstein-Hilbert action with respect to the dynamical field variable - the metric - when an additional term, proportional to the second derivatives of the dynamical variable, is formed 
\begin{equation}
    (2\kappa) \delta \Ssc_{\text{EH}} = \int_{\mathcal{M}}^{} d^4x \sqrt{-g} \stackrel{\circ}{G}_{\mu \nu} \delta g^{\mu \nu} + \int_{\mathcal{M}}^{} d^4x \sqrt{-g} g^{\mu \nu} \delta \Rc_{\mu \nu}.
\end{equation}
The first term, proportional to the Einstein tensor $\stackrel{\circ}{G}_{\mu \nu}$
\begin{equation}
    \stackrel{\circ}{G}_{\mu \nu} :=\Rc_{\mu \nu}-\frac{1}{2}g_{\mu \nu}\Rc,
\end{equation}
 recovers the correct left-hand side of the gravitational field equations. Let us, however, focus of the second part of the above equation. This other term - proportional to the variations of Ricci tensor $\delta \Rc_{\mu \nu}$ - is then, according to the standard procedure, firstly rewritten using the Palatini identity
\begin{equation}
    \delta \Rc_{\mu \nu} \equiv \nabla_\sigma \left(\delta \Gammac^\sigma{}_{\mu \nu} \right) - \nabla_\nu \left(\delta \Gammac^\sigma{}_{\mu \sigma} \right),
\end{equation}    
into a covariant gradient of a vector $\delta V^\mu{}$
\begin{equation}
    g^{\mu \nu}\delta \Rc_{\mu \nu}= \delta V^\mu{}_{;\mu}, \quad \delta V^\mu=g^{\rho \sigma} \delta \Gammac^\mu{}_{\rho \sigma}-g^{\rho \mu} \delta \Gammac^\sigma{}_{\rho \sigma}.
\end{equation}
Further comment is in place with regards to the tensorial nature of the vector $\delta V^\mu$, or rather, its constituents $\delta \Gammac^\mu{}_{\rho \sigma}$. As we have argued in the first chapter, Christoffel symbols $\Gammac^\mu{}_{\rho \sigma}$ transform inhomogeneously under diffeomorphisms. On the other hand, if we assume that  $\Gammac^\mu{}_{\rho \sigma}A^\rho dx^\sigma$ is a change in a vector $A=A^\rho\partial_\rho$ under parallel transport between two infinitesimally located points, it follows that  $\delta \Gammac^\mu{}_{\rho \sigma}A^\rho dx^\sigma$ describes the difference between two vectors obtained by two parallel transports - first, the vector $A$ is transported via the connection $\Gammac^\mu{}_{\rho \sigma}$, then via the connection $\left( \Gammac^\mu{}_{\rho \sigma}+ \delta \Gammac^\mu{}_{\rho \sigma}\right)$. Now, the difference between two vectors at the same point is yet again a vector. Therefore we reach the conclusion that the variation of connection $\delta \Gammac^\mu{}_{\rho \sigma}$ does indeed constitute a tensor.

Next, we shall make use of the Stoke's theorem in the following sense
\begin{equation}
    \int_{\mathcal{M}}^{} d^4x \sqrt{-g} A^\mu{};_\mu = \int_{\mathcal{M}}^{} d^4x \partial_\mu \left( \sqrt{-g}A^\mu \right)=\int_{\mathcal{\partial M}}^{} \epsilon d^3y \sqrt{|\gamma|} n_\mu A^\mu,
\end{equation}
where we have used the relation 
\begin{equation}
    A^\mu;_\mu=\frac{1}{\sqrt{-g}}\partial_\mu(\sqrt{-g}A^\mu).
\end{equation}
 All we need now is to identify $A^\mu$ with $\delta V^\mu$. But before we do so, let us briefly introduce the reader to the defining relations regarding the hypersurfaces. We start with the definition of a hypersurface as
 \begin{equation}
     f(x^\mu)=C,
 \end{equation}
 where $C$ is some constant parameter characterising the corresponding hypersurface, i.e. in spherical coordinates, we may define the boundary of constant radial coordinate simply as $r=r_c$. Next we define the normal vector $n^\mu$ to the hypersurface in the direction of increasing function $f: n^\mu f_{, \mu}>0$. In covariant language, the prescription is as follows
 \begin{equation}
     n_\mu = \frac{\epsilon f_{, \mu}}{(\epsilon g^{\mu \nu}f_{, \mu} f_{, \nu})^\frac{1}{2}}.
 \end{equation}
 Now, if we set the coordinates on the boundary as $\{ y^i\}$, with index $i=\{1,2,3\}$, it follows that the three vectors\footnote{By the abuse of notation, we chose the same symbol $e$ as for the general frames in the first section~\ref{chaptetradformalism}. The distinctive feature lies in indeces: for projectors, we adopt the induced 3-metric notation $i,j...=(1,2,3)$.} 
\begin{equation}\label{projectors}
    e^\mu{}_i=\left( \frac{\partial x^\mu}{\partial y^i} \right)_{\partial \mathcal{M}},
\end{equation}
are necessarily tangential to the hypersurface
\begin{equation}\label{tangentiality}
    n_\mu e^\mu{}_i=0.
\end{equation}
The induced metric $\gamma_{i j}$ is thus
\begin{equation}
    \gamma_{i j}=g_{\mu \nu} e^\mu{}_i e^\nu{}_j.
\end{equation}
At the same time, let us introduce the transverse metric
 \begin{equation}
     \gamma_{\mu \nu}=g_{\mu \nu}+n_\mu n_\nu,
 \end{equation}
 which essentially isolates the part of the metric that is transverse to the normal $n^\mu$. However, we need to keep in mind that the variation of the metric is fixed on the boundary $\delta g_{\mu \nu}|_{\partial \mathcal{M}}=0$, which implies that also the induced metric on the boundary is held fixed during the variation.

 Ultimately, we arrive at the final expression for the variation of the EH action
\begin{equation}\label{ehvar}
    (2\kappa) \delta \Ssc_{\text{EH}} = \int_{\mathcal{M}}^{} d^4x \sqrt{-g} \stackrel{\circ}{G}_{\mu \nu} \delta g^{\mu \nu}-\int_{\mathcal{\partial M}}^{} \epsilon d^3y \sqrt{|\gamma|} \gamma^{\rho \sigma} n^\mu \delta g_{\rho \sigma}{}_{,\mu}.
\end{equation}
The only problem here is that in order for us to kill this boundary term in the case of non-trivial boundary, one would need to impose the vanishing of normal derivates of the metric variations at that boundary $n^\mu \delta g_{\rho \sigma}{}_{,\mu}|_{\partial \mathcal{M}}=0$ along the already assumed condition of $\delta g_{\mu \nu}|_{\partial \mathcal{M}}=0$. But this would be no different than us resorting to both Dirichlet and Neumann boundary conditions at the same time. In other words, to make the variation self-consistent in the general case, one needs to impose the following condition \cite{Frolov}
\begin{equation}
    \left[ (\alpha - 1)\delta g_{\mu \nu} + \alpha\, n^\mu \delta g_{\rho \sigma}{}_{,\mu}\, \right]_{\partial \mathcal{M}} = 0.
\end{equation}
For $\alpha=0$, variational problem is well-defined for the Dirichlet boundary conditions $\delta g_{\mu \nu}|_{\partial \mathcal{M}}=0$. For $\alpha=1$ the variational problem is also well-defined, but for the Neumann boundary conditions $n^\mu \delta g_{\rho \sigma}{}_{,\mu}|_{\partial \mathcal{M}}=0$.

To amend this issue, the GHY term is constructed in such a way that it precisely compensates for this other boundary term, leaving the variation well-posed in the process, which is apparent after considering the variation of the trace of extrinsic curvature
\begin{equation}
    \delta \Kc = \frac{1}{2} \gamma^{\rho \sigma} n^\mu \delta g_{\rho \sigma}{}_{,\mu}.
\end{equation}
All in all, the variation of the total action is given by
\begin{equation}
      \delta \Ssc_{\text{tot}}=\delta \Ssc_{\text{EH}}+\delta \Ssc_{\text{GHY}}=\frac{1}{2\kappa}\int_{\mathcal{M}}^{} d^4x \sqrt{-g} \stackrel{\circ}{G}_{\mu \nu} \delta g^{\mu \nu}.
\end{equation}

\section{Renormalised action in GR}

If were to naively evaluate the total action \eqref{naiveghy} in the case of an asymptotically flat spacetime, it turns out that the gravitational action suffers from IR divergences. Therefore, to regularise the action in such a case, one also needs to subtract the corresponding boundary term proportional to the trace of extrinsic curvature of the flat spacetime $\Kc_0$. In practice, this usually boils down to subtracting the GHY boundary term for the flat Minkowski metric. Albeit such a spacetime does not have a natural boundary, one can nevertheless set a hypersurface of a constant radial coordinate $r=r_c$ and send $r_c$ asymptotically to infinity, as we shall demonstrate shortly. Ultimately, we get the following regularised gravitational action for a spacetime with a boundary \cite{Gibbons:1976ue,Ortín_2004} 
\begin{equation}\label{totalrenormghyaction}
    \Ssc_{\text{tot}}=\Ssc_{\text{EH}}+\Ssc_{\text{GHY,0}}=-\frac{1}{2\kappa}\int_{\mathcal{M}}^{} d^4x \sqrt{-g} \Rc - \frac{1}{\kappa}\int_{\partial \mathcal{M}}^{} d^3y \epsilon \sqrt{|\gamma|} \left(\Kc - \Kc_0\right).
\end{equation}

To illustrate the need for the background subtraction and at the same time to present the reader with the GH method in practice, let us turn to the case of a flat Minkowski metric in the spherical coordinates
\begin{equation}
    \eta_{\mu \nu} =  
    \begin{pmatrix}
        1& 0 & 0 & 0\\
        0 & -1 & 0 & 0\\
        0 & 0 & -r^2 & 0\\
        0 & 0 & 0 & -r^2 \sin^2 \theta
    \end{pmatrix},
\end{equation}
with the inverse metric 
\begin{equation}
    \eta^{\mu \nu} =  
    \begin{pmatrix}
        1& 0 & 0 & 0\\
        0 & -1 & 0 & 0\\
        0 & 0 & -\frac{1}{r^2} & 0\\
        0 & 0 & 0 & -\frac{\csc^2{\theta}}{r^2}
    \end{pmatrix}.
\end{equation}
It is now a trivial task to show that the EH action is, in this case, equal to zero. All we need to realise is that the Minkowski metric, when expressed in Cartesian coordinates, assumes a rather simple form, with the components being constants
\begin{equation}
    \eta_{\mu \nu} = 
    \begin{pmatrix}
        1& 0 & 0 & 0\\
        0 & -1 & 0 & 0\\
        0 & 0 & -1 & 0\\
        0 & 0 & 0 & -1
    \end{pmatrix}.
\end{equation}
Hence, the Christoffel symbols \eqref{RLCchrist} are found to vanish. This in turn leads to vanishing Riemann curvature tensor \eqref{riemanncurvature}, and therefore the Ricci scalar \eqref{riccisc} vanishes as well. Now, thanks to the tensors expressed in coordinate basis being covariant with respect to coordinate transformations, vanishing of Ricci scalar in one coordinate system ensures that it vanishes in any other system.

The non-trivial part comes when we try to evaluate the action assigned to the GHY boundary term. Despite us knowing the result in advance - as the Minkowski metric represents a flat spacetime, and thus trivially resolves the field equations of general ralativity in vacuum - and as such, we expect the total gravitational action to vanish, let us carry on with the computation along the standard procedure. To this end, we firstly ought to specify the boundary along which we evaluate the GHY term. This is when the spherical coordinates become particularly convenient, as we can now set the hypersurface of a constant radial coordinate simply as $r=r_c$, with the constant $r_c$ sent to infinity. Next we determine the unit normal 
\begin{equation}
    n_\mu \propto \partial_\mu (r-r_c)=\delta_{\mu r}.
\end{equation}
As the normal vector for this particular hypersurface is spacelike, the normalisation is defined as
\begin{equation}
    n_\mu n^\mu = \epsilon \equiv -1.
\end{equation}
Lastly, we need to account for the correct sign of the normal vector such that it is outward-pointing. All in all, we get the following
\begin{equation}
    n_\mu=-\delta_{\mu r} \equiv (0, -1, 0, 0), \quad n^\mu=g^{\mu \nu}n_\nu=-g^{rr}=\delta^{\mu r}.
\end{equation}
The trace if the extrinsic curvature is specified as
\[
    \Kc = n^\mu;_\mu=\frac{1}{\sqrt{-g}}\partial_\mu(\sqrt{-g}n^\mu)=
\]
\begin{equation}
    = \frac{1}{r^2 \sin{\theta}}\partial_\mu(r^2 \sin{\theta} \delta^{\mu r})= \frac{2}{r}.
\end{equation}
The $\sqrt{|\gamma|}$ for the metric $\gamma_{i j}$ induced on the boundary is found straightforwardly as 
\begin{equation}
    \sqrt{|\gamma|}=r_c^2 \sin{\theta}
\end{equation}
If we were to naively evaluate the GHY action alone, we would find out that it is, indeed, rather badly divergent after taking the limit of $r_c \rightarrow \infty$
\[
    \Ssc_{\text{GHY}} = \frac{1}{\kappa}\int_{}^{} d^3y  \sqrt{|\gamma|} \Kc |_{r=r_c} = \frac{2}{\kappa}\int_{}^{}  dt\, d\Omega\, r_c  =
\]
\begin{equation}
    = \int_{}^{}  dt\, r_c \longrightarrow \infty.
\end{equation}
To obtain a finite, albeit trivial action, one thus needs to subtract the asymptotically flat background - which is, in this case, the same Minkowski metric we used for the underlying spacetime. The regulating term is found analogously as
\begin{equation}
    \Kc_0=\frac{2}{r},
\end{equation}
and thus we ultimately obtain the desired vanishing total action
\begin{equation}
    \Ssc_{\text{tot}}=\Ssc_{\text{EH}}+\Ssc_{\text{GHY,0}}=\frac{1}{2\kappa}\int_{\mathcal{M}}^{} d^4x \sqrt{-g} \Rc + \frac{1}{\kappa}\int_{\partial \mathcal{M}}^{} d^3y \epsilon \sqrt{|\gamma|} \left(\Kc - \Kc_0\right) = 0.
\end{equation}

\section{TEGR action}
We start with the introduction of the teleparallel Lagrangian \cite{Aldrovandi:2013wha, Krssak:2018ywd}
\begin{equation}\label{tellag}
    \Lb_{\text{TG}}=\frac{h}{2\kappa}(\frac{1}{4} \Tb^\rho{}_{\mu \nu} \Tb_\rho{}^{\mu \nu}+\frac{1}{2} \Tb^\rho{}_{\mu \nu} \Tb^{\nu \mu}{}_\rho - \Tb^\rho{}_{\mu \rho} \Tb^{\nu \mu}{}_\nu) \equiv \frac{h}{2\kappa} \Tb,
\end{equation}
with $\Tb^\rho{}_{\mu \nu}= {h_a}^\rho \Tb^a{}_{\mu \nu}$ and $h\equiv \text{det}({h^a}_\mu)$. From now on we shall collectively refer to this particular combination - quadratic in torsion tensors - as a torsion scalar $\Tb$, in complete analogy to Einstein-Hilbert action and Ricci scalar $\Rc$. As each term of the teleparallel Lagrangian is constructed from the torsion tensor, and the contractions preserve "tensoriality", the ensuing torsion scalar turns out invariant under both global coordinate and local Lorentz transformations. A crucial property of this Lagrangian lies in its equivalence up to a surface term to the standard Einstein-Hilbert action
\begin{equation}
    \Lb_{\text{TG}}=\Lc_{\text{EH}}-\partial_\mu \left( \frac{h}{\kappa}\Tb^\mu \right),
\end{equation}
with $\Tb^\mu=\Tb^{\nu \mu}{}_\nu$ being the torsion vector. As the surface term does not contribute to the equations of motion upon variation of the action, the dynamical content of both theories must be the same, hence the name Teleparallel Equivalent of General Relativity (TEGR). Nevertheless, to fully establish the equivalence, notice that the resulting Euler-Lagrange expression after variation of the TG action has to be symmetric. This is achieved exclusively for the combination of constant parameters as is shown in \eqref{tellag}.

Variation of the teleparallel gravitational action $\mathcal{L}=\Lb_{\text{TG}}+\mathcal{L}_{\text{source}}$ with respect to the tetrad field ${h^a}_\mu$ leads to the following Euler-Lagrange expression
\begin{equation}
        {E_a}^\mu \equiv \partial_\sigma (h \Sb_a{}^{\mu \sigma})-\kappa h \Jb_a{}^\mu = \kappa h {\Theta_a}^\mu,
    \end{equation}
where we have defined the superpotential $\Sb_a{}^{\mu \sigma}$ as
\begin{equation}
    \Sb_a{}^{\mu \sigma}=\frac{1}{2}(\Tb^{\sigma \mu}{}_a+\Tb_a{}^{\mu \sigma}-\Tb^{\mu \sigma}{}_a)-{h_a}^\sigma \Tb^{\theta \mu}{}_\theta+{h_a}^\mu \Tb^{\theta \sigma}{}_\theta,
\end{equation}
and the gauge current $\Jb_a{}^\mu$
\begin{equation}
    \Jb_a{}^\mu=\frac{1}{\kappa} {h_a}^\lambda \Sb_c{}^{\nu \mu} \Tb^c{}_{\nu \lambda}-\frac{{h_a}^\mu}{h}\Lb+\frac{1}{\kappa}\omegab^c{}_{a \sigma} \Sb_c{}^{\mu \sigma},
\end{equation}
with the right-hand side constituting an energy-momentum tensor ${\Theta_a}^\mu$ of the matter source field defined as a variation of the matter Lagrangian $\mathcal{L}_{\text{source}}$ with respect to the tetrad
\begin{equation}
    h \Theta^a{}_\mu=\frac{\delta \mathcal{L}_{\text{source}}}{\delta h_a{}^\mu}.
\end{equation}

Incidentally, we may want to re-express the equations of motion strictly in terms of spacetime indices as follows
\begin{equation}
        {E_\mu}^\rho \equiv \partial_\sigma (h \Sb_\mu{}^{\rho \sigma})+\kappa h \tb_\mu{}^\rho =  \kappa h {\Theta_\mu}^\rho,
\end{equation}
where we have introduced the energy-momentum pseudotensor $\tb_\mu{}^\rho$ as
\begin{equation}
      \tb_\mu{}^\rho = \frac{1}{\kappa} \Gammab^\alpha{}_{\sigma \mu} \Sb_\alpha{}^{\sigma \rho}+\frac{1}{h}{\delta_\mu}^\rho \Lb, 
\end{equation}
with $\Gammab^\alpha{}_{\sigma \mu}$ being the teleparallel connection coefficients \eqref{conneccoeftg}.

 A subtle, but nonetheless important observation is that the variation of the TG Lagrangian with respect to the spin connection $\omegab^a{}_{b \mu}$ is trivially satisfied, which is apparent from the following decomposition \cite{Krssak:2015lba}
\begin{equation}
    \Lb (h^a{}_{\mu},\omegab^a{}_{b \mu})=\Lb (h^a{}_{\mu},0) + \frac{1}{\kappa}\partial_\mu(h \omegab^\mu).
\end{equation}
A natural question immediately arises: Why do we need then to specify the teleparallel spin connection, if it enters the Lagrangian through the surface term only? It is certainly true that upon variation of the action with respect to the tetrad (the dynamical field variable of TG), we may just as well let the spin connection vanish, and the dynamical content of ensuing equations of motions is left unchanged. Evidently, as the variation with respect to the spin connection is trivial, there are simply no extra equations governing its own dynamics. Consequently, just as the tetrad, the spin connection is also left determined up to a local Lorentz transformation, justifying the choice of a vanishing spin connection for the sake of simplicity. On the other hand, it also means that we can utilise this freedom to some other end. Namely, we may want to regularise the typically divergent action, as we are about to see in the following sections.

\section{Spin connection associated to a given tetrad}

Here we return to the concept of inertial frames and their associated spin connection, for we have yet to present the reader with some sort of a procedure of how to specify its form, as the naive assumption of a vanishing spin connection ultimately causes an unwanted divergent behaviour of the teleparallel action.

As has already been hinted at along the eq. \eqref{inertialframe}, the "inertial"\footnote{Although this particular wording follows the conventions in \cite{Aldrovandi:2013wha}, and naturally stems from the use of the inertial frames, we shall mostly try to refrain from it as it becomes ambiguous in the presence of gravity.} spin connection arises in Special Relativity in a particular class of inertial frames, for which the coefficients of anholonomy vanish ${f'^c}_{a b}=0$. This in turn leads to the holonomic (coordinate) expression of the tetrad field
\begin{equation}
    {h'^a}_\mu=\partial_\mu x'^a,
\end{equation}
with $x'^a \equiv x'^a(x^\mu)$ being the tangent space coordinates, a 4-vector. That is, it transforms via the local Lorentz transformation $\Lambda^a{}_b (x)$ as 
\begin{equation}
    x^a=\Lambda^a{}_b(x) x'^b.
\end{equation}
A general frame ${h^a}_\mu$ is thus obtained by demanding the following covariant transformation law
\begin{equation}
    {h^a}_\mu=\Lambda^a{}_b(x){h'^b}_\mu.
\end{equation}
Utilising the above relations, a straightforward computation shows
\begin{equation}
    {h^a}_\mu=\partial_\mu x^a + \Lambda^a{}_c \partial_\mu \Lambda^c{}_b \equiv \partial_\mu x^a + \omegab^a{}_{b \mu} x^b, 
\end{equation}
where we have defined the spin connection as
\begin{equation}\label{tgspincon}
    \omegab^a{}_{b \mu} = \Lambda^a{}_c \partial_\mu \Lambda_b{}^c.
\end{equation}
Recalling now the transformation property of a general spin connection ${\omegag^a}_{b \mu}$, we see that the spin connection $\omegab^a{}_{b \mu}$ in \eqref{tgspincon} is but a connection obtained from a Lorentz transformation of the vanishing spin connection $\omegab'^c{}_{d \mu}=0$
\begin{equation}
    \omegab^a{}_{b \mu}=\Lambda^a{}_c \omegab'^c{}_{d \mu} \Lambda_b{}^d + \Lambda^a{}_c \partial_\mu \Lambda_b{}^c \equiv \Lambda^a{}_c \partial_\mu \Lambda_b{}^c.
\end{equation}
This sometimes leads people to say that the spin connection $\overset{\bullet}{\omega^a}_{b \mu}$ represents the inertial effects present in the new, Lorentz transformed frame ${h^a}_\mu$, as we are reduced to the trivial case of special relativity, without any gravitational effects being present. 

\bigskip

Generalising now this concept of the spin connection $\omegab^a{}_{b \mu}$ to the teleparallel gravity, let us define a unique class of frames, coined as the proper frames, characterised by a vanishing spin connection $\{ {\tilde{h}^a}_\mu,0 \}$. Any other Lorentz transformed frame will thus necessarily lead to a non-vanishing spin connection $\{ {h^a}_\mu,\omegab^a{}_{b \mu}\}$ of the form $\Lambda^a{}_c \partial_\mu \Lambda_b{}^c$. Nonetheless, it is not a priori clear how one arrives at such a spin connection associated to a given tetrad ${h^a}_\mu$, due to one aspect being dissimilar to the case of trivial tetrads: the presence of gravity. This becomes apparent after considering the case of standard general relativity, where the sole presence of the gravitating matter curves the spacetime and thus modifies the metric. In TEGR, this directly translates to a modification of the tetrad, despite us interpreting the gravitational effects on the spacetime geometry in terms of torsion instead of curvature. On the other hand, if we were somehow able to let the effects of gravity terminate, it would suddenly make perfect sense to talk in terms of "inertial" spin connections. To this end, we formally define a "reference" tetrad $h^a_{(r) \mu}$ - in which the gravity is "switched off" - by sending the gravitational constant G to zero (or more rigorously, by assuming the asymptotic flatness) \cite{Lucas:2009nq,Krssak:2015lba,Krssak:2015rqa} 
\begin{equation}
    h^a_{(r) \mu} \equiv {h^a}_\mu |_{G \rightarrow 0}.
\end{equation}
We may thus treat such a reference tetrad $h^a_{(r) \mu}$ (formally) as the trivial case discussed at the beginning of this section. Realising that for such a case, the torsion tensor in non-coordinate basis $\Tb^a{}_{b c}(h^a_{(r) \mu},\omegab^a{}_{b \mu})$ vanishes everywhere
\begin{equation}
    \Tb^a{}_{b c}(h^a_{(r) \mu},\omegab^a{}_{b \mu})=\omegab^a{}_{c b}-\omegab^a{}_{b c}-{f^a}_{bc}(h_{(r)})=0,
\end{equation}
which yields the relation for the inertial spin connection associated with the reference tetrad $h^a_{(r) \mu}$ as 
\begin{equation}\label{reftetspincon}
    \omegab^a{}_{b \mu}=\frac{1}{2} h^c_{(r) \mu} [{{f_b}^a}_c(h_{(r)})+{{f_c}^a}_b(h_{(r)})-{f^a}_{b c}(h_{(r)})].
\end{equation}
 Moreover, as the reference tetrad $h^a_{(r) \mu}$ differs from the "full" tetrad ${h^a}_\mu$ only in the gravitational content (the inertial content of both tetrads being equal by construction), the spin connection $\omegab^a{}_{b \mu}$ is also a spin connection associated with the full tetrad ${h^a}_\mu$. Thus in terms of the inertial effects, the teleparallel spin connection of the full tetrad coincides with the Levi-Civita spin connection with respect to the reference tetrad, which becomes apparent after considering the connection decomposition \eqref{telspincondec}
 \begin{equation}
    \omegab^a{}_{b \mu} (h^a_{(r) \mu}) = \omegac^a{}_{b \mu} (h^a_{(r) \mu})+ \Kb^a{}_{b \mu} (h^a_{(r) \mu}) \equiv \omegac^a{}_{b \mu} (h^a_{(r) \mu}).
 \end{equation}
 As the reference tetrad solves for the vanishing torsion tensor, the contortion vanishes as well. Taking then into account the above argumentation, we get the following generalised relation
 \begin{equation}
     \omegab^a{}_{b \mu} (h^a{}_\mu) \equiv  \omegac^a{}_{b \mu} (h^a_{(r) \mu}).
 \end{equation}
While it is true that the spin connection $\omegab^a{}_{b \mu}$ is derived via vanishing of torsion when understood as a function of the reference tetrad $h^a_{(r) \mu}$, this does not necessarily violate the condition of non-vanishing torsion of the full tetrad $h^a{}_\mu$ in the definition of teleparallel connection. This is apparent after recalling the formula for torsion with mixed components \eqref{torsion}: both the tetrad and the spin connection enter there. Setting this expression equal to zero for some auxiliary, reference tetrad is to be understood as a tool that merely allows one to establish a relation between the two objects.

We have thus arrived at the spin connection associated to the original tetrad ${h^a}_\mu$ by determining the inertial effects as if it were an inertial tetrad in the absence of gravity; and then finding such a connection that precisely compensates for these effects. The key consequence is that the tensors constructed from the given tetrad and its associated spin connection are proper tensors in the sense that they transform covariantly under simultaneous local Lorentz transformation of both the tetrad and the spin connection. 

\section{Renormalised TEGR action}

If we were to naively associate vanishing spin connection to a given tetrad when aiming to evaluate the teleparallel action, the presence of inertial effects - which typically do not vanish at infinity - causes the action to diverge
\begin{equation}
     \Ssb (h^a_{(r) \mu},0)=\int_{\mathcal{M}}^{} \Lb(h^a_{(r) \mu},0) \longrightarrow \infty.
\end{equation}
Recalling now that the torsion as a function of both the reference tetrad and the associated spin connection vanishes $\Tb^a{}_{\mu \nu}(h^a_{(r) \mu},\omegab^a{}_{b \mu}) \equiv 0$, we straightforwardly get
\begin{equation}
     \Ssb(h^a_{(r) \mu},\omegab^a{}_{b \mu})=0.
\end{equation}
The following line of reasoning ought to be already familiar: as the action of the reference tetrad represents the inertial effects due to gravity being switched-off, by considering the appropriate spin connection, we remove all the spurious inertial contributions, naturally leading to a vanishing action. As we have already established the equivalence between the reference and full tetrad in what concerns the inertiality, the action of the full tetrad consequently represents the gravitational effects only. Assuming the divergences are caused by the inertial effects, such an action with its associated connection is finite 
\begin{equation}\label{tgrenaction}
      \Ssb_{\text{ren}}=\int_{\mathcal{M}}^{} \Lb(h^a_{\mu},\omegab^a{}_{b \mu}) \nrightarrow \infty.
\end{equation}
Therefore, by considering the appropriate spin connection, we arrive at a renormalised teleparallel action, as it becomes free of typically divergent inertial effects. This has profound consequences for the validity of various conserved quantities, gravitational energy-momentum, or in our case, when we are evaluating the action of a given theory.

To demonstrate this point, let us yet again reduce to the case of Minkowski spacetime in spherical coordinates 
\begin{equation}
    \eta_{\mu \nu}=\text{diag}(1, -1, -r^2, -r^2 \sin^2{\theta}),
\end{equation}
and consider the following diagonal tetrad
\begin{equation}\label{diagminktet}
    {h^a}_\mu = \text{diag}(1,1,r,r \sin \theta), \quad h\equiv \text{det}(h^a{}_\mu)=r^2 \sin{\theta},
\end{equation}
such that
\begin{equation}
    \eta_{\mu \nu}=\eta_{a b}h^a{}_\mu h^b{}_\nu,
\end{equation}
for if we were to stay in Cartesian coordinates, the torsion tensor, and hence the teleparallel Lagrangian \eqref{tellag} as functions of zero spin connection would vanish trivially. In the spherical coordinates, however, we need to account for the inertial effects. Indeed, if we calculate the torsion scalar as a function of tetrad \eqref{diagminktet} and assuming the vanishing spin connection
\begin{equation}
    T(h^a{}_\mu,0)=\frac{2}{r^2},
\end{equation}
and evaluate the teleparallel action 
\begin{equation}
    \Ssb (h^a{}_\mu,0)= \frac{1}{2 \kappa}\int_{\mathcal{M}}^{} d^4x\, h\, T(h^a_{ \mu},0)=\frac{1}{ \kappa}\int_{\mathcal{M}}^{} d^4x\, \sin{\theta} = \frac{1}{2} \int_{}^{} dt \, r,
\end{equation}
we see that after the limit $r \rightarrow \infty$, the action diverges\footnote{It also trivially holds that the action diverges with respect to the time integral, as the temporal coordinate is typically unbounded. We thus have several options already: either to integrate over some compact interval $t\in [t_1, t_2]$, or we do not evaluate the time integral at all. Lastly, we may define the so-called action growth $\frac{d \mathcal{S}}{dt}$ as in \cite{Brown:2015bva, Brown:2015lvg}, but have opted to remain a bit more conservative with the use of such novel ideas.}.

On the other hand, if we utilise the method of reference tetrad and specify the corresponding spin connection \eqref{reftetspincon}, with the non-zero components given as
\begin{equation}
    \omegab^1{}_{2 \theta}=-1, \quad  \omegab^1{}_{3 \phi}=-\sin\theta, \quad  \omegab^2{}_{3 \phi}=-\cos\theta,
\end{equation}
we find that the torsion scalar, now as a function of both tetrad and spin connection, vanishes altogether 
\begin{equation}
    T(h^a{}_\mu, \omega^a{}_{b \mu})= 0.
\end{equation}
The action thus becomes desirably regularised
\begin{equation}
    \Ssb(h^a{}_\mu, \omega^a{}_{b \mu})=\frac{1}{2 \kappa}\int_{\mathcal{M}}^{} d^4x\, h\, T(h^a{}_\mu, \omega^a{}_{b \mu}) = 0. 
\end{equation}

\chapter{Action of a Schwarzschild black hole\label{bhaction}}
Here we shall study the spherically symmetric, static and asymptotically flat vacuum solution in the theory of gravity. First, through the lenses of the well-established method of GHY boundary terms within the standard GR to obtain a prediction for the Schwarzschild black hole action. Subsequently, we compare this result with the teleparallel framework. Lastly, we employ an alternative approach by considering the idea of a canonical frame, only to hint at the hidden inconsistencies in the (ab)use of Stokes' theorem. 

\section{Schwarzschild in GR}

Let us now have a look at the evaluation of the action for the Schwarzschild solution within the familiar bounds of general relativity. As has already been revealed in section~\ref{chapghy}, an addition of a certain boundary term - also known as the GHY boundary term - must be considered. As it turns out, this other term has to be employed for each non-trivial boundary of the spacetime manifold in order for us to obtain a finite, non-zero result all while keeping the variations on the boundary well-defined.

\subsection{Singular behaviour of the event horizon}

Before we apply this whole procedure to the Schwarzschild solution, we need to first discuss the problem of the coordinate singularity at the black hole horizon. This coordinate singularity at $r_s=2M$ divides the Schwarzschild coordinates $(t,r,\theta, \phi)$ in two disconnected patches, which we aptly call the interior region $0\leq r < r_s$, and the exterior region $r>r_s$. If we want to evaluate the action across the whole black hole, we better do something about this singularity, as we can integrate over a smooth manifold only. Fortunately, it turns out that the exterior Schwarzschild solution allows for an analytical continuation to the Euclidean one. In doing so, we however loose the ability to say anything about the interior region. We will get back to this issue later in this section.

 We start by performing the Wick's rotation $t \rightarrow \tau=it$ of the standard Schwarzschild metric
 \begin{equation}\label{schwmetric}
     ds^2= f(r) dt^2 - f(r)^{-1} dr^2 - r^2 d\Omega^2_{(2)}, \quad f(r) \equiv 1-\frac{2M}{r},
 \end{equation}
 to obtain the Euclidean version of Einstein-Hilbert action \cite{Horowitz:1992jp}, characterised by the Euclidean Schwarzschild metric, now with the signature of $(-,-,-,-)$
\begin{equation}
    ds^2=- f(r) d\tau^2 - f(r)^{-1} dr^2 - r^2 d\Omega^2_{(2)}.
\end{equation}
Let us now focus on the $r - \tau$ plane part of the metric 
\begin{equation}
    -d\sigma ^2 \equiv f(r) d\tau^2 + f(r)^{-1}dr^2.
\end{equation}
Assuming that there is a non-degenerate event horizon at $r=r_s$, it follows that near this horizon $f(r) \approx f'(r_s) \xi$ with $\xi = r-r_s$. Next let us define $\rho = 2\sqrt{\xi/f'(r_s)}$ so that the metric is brought to a more familiar form
\begin{equation}
    -d\sigma^2=\frac{f'(r_s)^2}{4}\rho^2 d\tau^2+d\rho^2 \equiv F(\rho)^2 d\tau^2 + d\rho^2,
\end{equation}
with $F(\rho) \equiv \frac{f'(r_s)}{2}\rho=\frac{1}{4M}\rho$ being a function which measures the circumference of circles ($\int_{}^{} \sqrt{g_{\rho \rho}} d\tau =\int_{}^{} F(\rho) d\tau = 2\pi F(\rho)$) and the coordinate $\rho$ representing the physical length of radial curves($\int_{}^{}\sqrt{g_{\rho \rho}}d\rho=\rho$); which we recognise as a 2-plane metric in polar coordinates if and only if $\tau' \in [0,2\pi]$. Otherwise, it represents metric of a cone with quasiregular (conical) singularity at $\rho=0$. Being quasiregular \cite{Ellis:1977pj}, there is no curvature obstacle to extension, which makes a perfect sense recalling that the singularity at the horizon is but a coordinate one (the Kretschmann invariant is regular at $r_s=2M$).

More precisely, let the angle $\alpha$ represent an angle of deviation of the conical surface from the axis of rotational symmetry. $\rho$ then necessarily measures the radial distance from the tip to a certain point on the surface of cone. Let us denote the perpendicular distance of that particular point on the surface from the main axis by $R$. 
\begin{figure}[!ht]
\centering
\includegraphics[width=.35\linewidth]{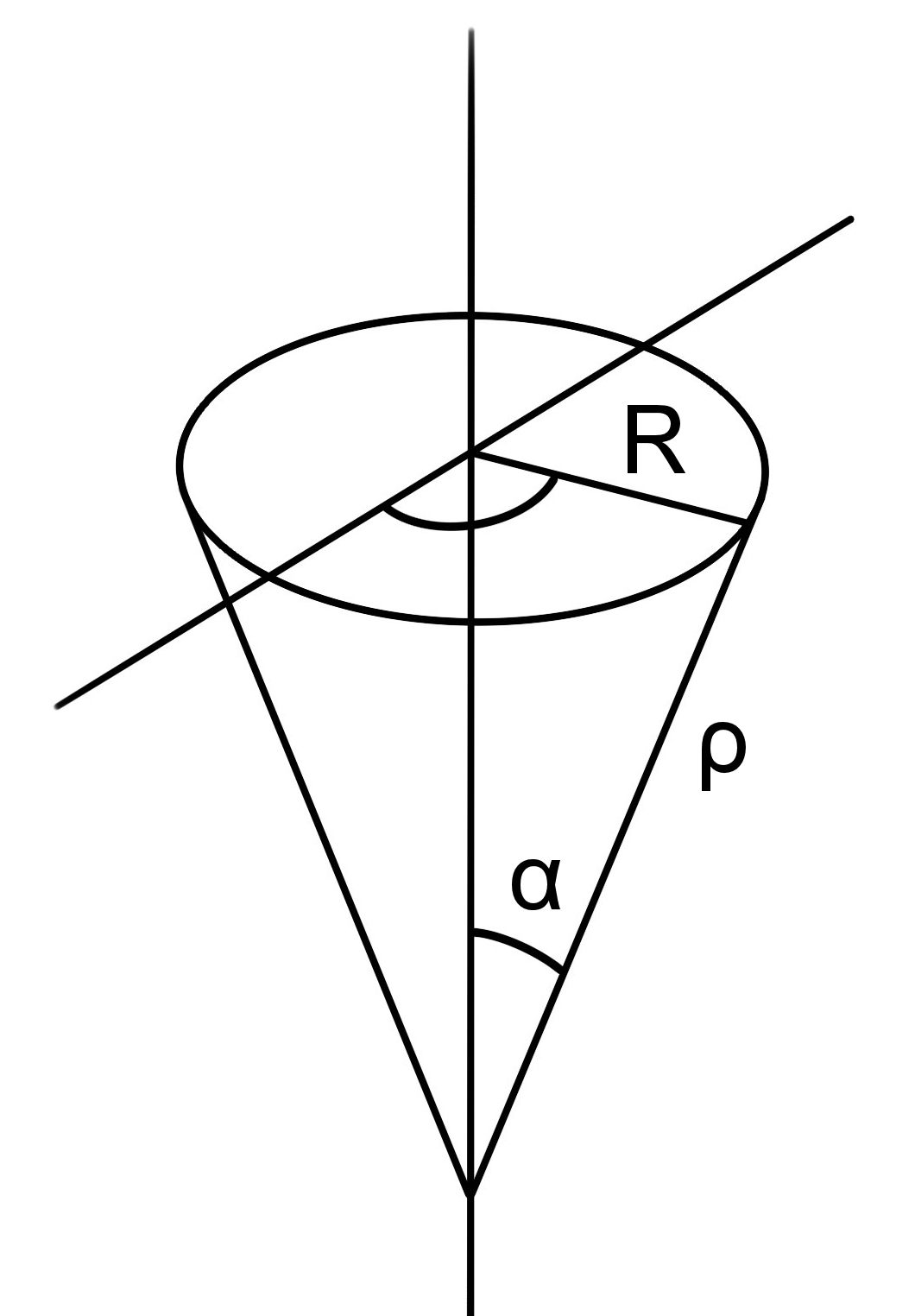}
\captionof{figure}{A cone displaying the respective coordinates.}
\end{figure}
The conical singularity is then defined as
\begin{equation}
    \lim_{\rho \to 0} \frac{2 \pi R}{2 \pi \rho}= \sin{\alpha} <1, \quad \forall \alpha \in \left[0,\frac{\pi}{2}\right).
\end{equation}
That being said, setting $\alpha$ equal to $\frac{\pi}{2}$ by hand, we would recover the ordinary flat 2-plane. There is, however, an ingenious way of how to get rid of this pesky conical singularity while preserving the conical structure. The correct question to ponder is what should the period be equal to for us to to set the above ratio to one?
\begin{equation}
    \int_{0}^{\beta} F(\rho) d\tau = \frac{\beta}{4M} \rho \equiv 2\pi R.
\end{equation}
In other words, we choose a period $\beta$ of the angular coordinate $\tau$ such that the radial distance $\rho$ coincides with the perpendicular distance from the main axis $R$, leading us to the value $\beta=8 \pi M$. The full metric of the Euclidean black hole
\begin{equation}
    -ds^2= -d\sigma^2 + r^2 d\Omega^2_{(2)},
\end{equation}
with such periodicity $\tau \in [0,\beta=8\pi M]$ is thus made to be regular at $\rho=0$, resembling a semi-infinite 'cigar' ($\times$ 2-sphere) with the topology $\mathbb{R}{}^2 \times S^2$; or rather in the sense of a periodic time coordinate $\tau$, the topology becomes $S^1 \times S^2$.
\begin{figure}[!ht]
\centering
\includegraphics[width=.6\linewidth]{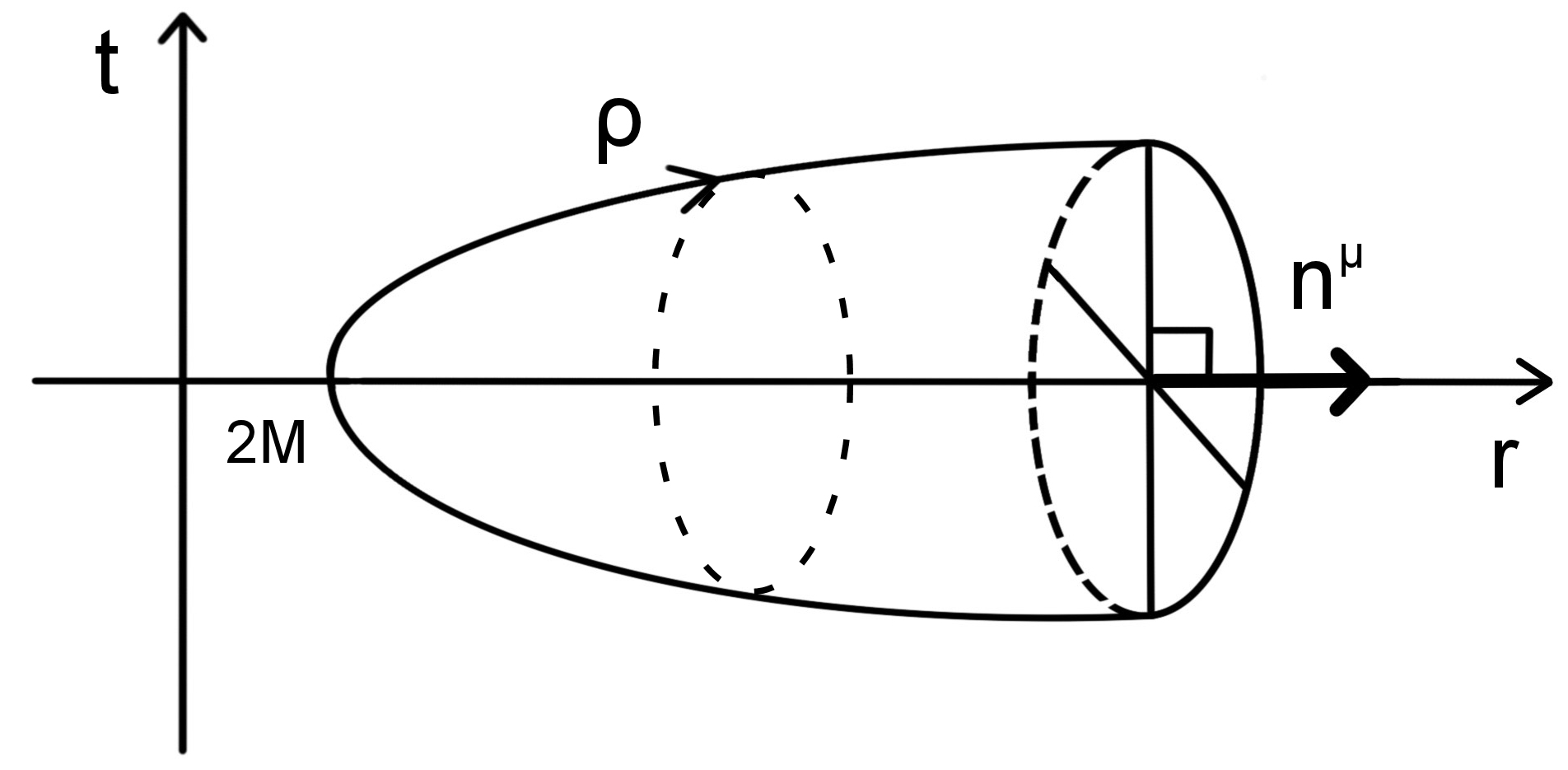}
\captionof{figure}{The semi-infinite cigar embedded in Minkowski spacetime with the normal vector $n^\mu$ to the hypersurface of constant $r$.}
\end{figure}

\subsection{Evaluation of the total action \label{chapghy}}

Now we are fully set to evaluate the action for the Euclidean Schwarzschild solution utilising the GHY boundary and regulating  terms \eqref{totalrenormghyaction}. With the conical singularity out of the way, the only possible choice of a boundary is in this case the hypersurface $r=r_c$, with the constant $r_c$ sent to infinity. Next we need to specify a vector normal to the aforementioned hypersurface $r-r_c=0$, which yields the proportionality $n_\mu \propto \partial_\mu (r-r_c)=\delta_{\mu r}$. Since $n_\mu n^\mu=\epsilon = -1$ (as it is spacelike), we need to make sure to get the sign right so that the normal is outward-pointing.  Finally, the proportionality constant is found by demanding the normalisation to unity
\begin{equation}
    n_\mu = - \frac{\delta_{\mu r}}{\sqrt{-n^2}}= -\sqrt{-g_{r r}}\delta_{\mu r} \equiv - \frac{\delta_{\mu r}}{\sqrt{f(r)}}.
\end{equation}
The trace of the extrinsic curvature $\Kc$ is found straightforwardly as
  \[  \Kc = n^\mu_{;\mu}=\frac{1}{\sqrt{-g}}\partial_\mu(\sqrt{-g}n^\mu)= \]
  \[ =\frac{1}{r^2 \sin{\theta}} \partial_\mu \left(r^2 \sin{\theta} \sqrt{f(r)} \delta^{\mu r}\right)= \]
  \[
  = \sqrt{1-\frac{2M}{r}} \left( \frac{M}{r^2-2Mr}+\frac{2}{r}\right)
  \]
\begin{equation}\label{extcurvschw}
    = \frac{1}{r^2\sqrt{f}} \left( -3M+2r \right),
\end{equation}
and the trace of the extrinsic curvature for the flat background $\Kc_0$ is then obtained after the limit as $r \rightarrow \infty$ in $g_{tt}$ and $g_{rr}$ part of the metric, resulting in
\begin{equation}\label{extcurvflatbcg}
    \Kc_0= \frac{2}{r}.
\end{equation}
Bringing all the pieces together gives us
\begin{equation}
    \left( \Kc - \Kc_0 \right) = \sqrt{1-\frac{2M}{r}} \left( \frac{M}{r^2-2Mr}+\frac{2}{r}\right)-\frac{2}{r} = \frac{-3M+2\left(1-\sqrt{f}\right)r}{r^2\sqrt{f}}.
\end{equation}
The spacetime 4-dimensional metric $g_{\mu \nu }$ induces the so-called transverse metric $\gamma_{\mu \nu}$ \cite{Padmanabhan:2010zzb}, which isolates the part of the metric that is transverse to the normal $n^\mu$
   \[ ds^2= \gamma_{\mu \nu} dx^\mu dx^\nu = (g_{\mu \nu} + n_\mu n_\nu)dx^\mu dx^\nu = \]
   \[
    = g_{\tau \tau}d\tau^2 + g_{r r}dr^2-r^2 d\Omega^2_{(2)} - g_{r r}dr^2 = \]
\begin{equation}
    = -f(r)d\tau^2-r^2 d\Omega^2_{(2)}.
\end{equation}
If we now utilise the parametric form $x^\mu=x^\mu(y^i)$ for the three-dimensional hypersurface within the four-dimensional spacetime, the ensuing vectors $e^\mu{}_i=(\partial x^\mu / \partial y^i)$ in \eqref{projectors} are then tangent to the curves within the hypersurface and orthogonal to the normal $n^\mu$ due to \eqref{tangentiality}. These relations then allows for the definition of the metric induced on the particular hypersurface $\gamma_{i j}$ by the following prescription
\begin{equation}
    \gamma_{i j}= \gamma_{\mu \nu} e^\mu{}_i e^\nu{}_j.
\end{equation}
In our case, the parameters are chosen as $y^i=(\tau,\theta, \phi)$, which straightforwardly yields the following form of the projectors $e^\mu{}_i$ as a 4x3 matrix
\begin{equation}
    e^\mu{}_i = 
    \begin{pmatrix}
        1 & 0 & 0 \\
        0 & 0 & 0 \\
        0 & 1 & 0 \\
        0 & 0 & 1
    \end{pmatrix},
\end{equation}
leading us to the desired form of the induced 3-metric on the hypersurface $r=r_c$
\begin{equation}
    \gamma_{i j }=
    \begin{pmatrix}
        1-\frac{2M}{r} & 0 & 0\\
        0 & -r^2 & 0\\
        0 & 0 & -r^2 \sin^2 \theta
    \end{pmatrix}.
\end{equation}
Consequently, we find that
\begin{equation}
    \sqrt{|\gamma|} =\sqrt{f(r_c)}\, r_c^2 \sin{\theta}.
\end{equation}
Now we finally move to the evaluation of the action. Recalling that the Euclidean Schwarzschild solution is found as a solution in vacuum $(\leftrightarrow \Rc_{\mu \nu}=0)$, it follows that the Einstein-Hilbert action vanishes. Hence all we need is to evaluate the following\footnote{By applying Wick's rotation, the action becomes purely imaginary, which is desired in the path integral approach and subsequent saddle-point approximation - however, it is not needed for our purposes here, and we continue without the explicit integration over the $\tau$ coordinate.}
   \[ \Ssc_{\text{GHY},0}=-\frac{1}{\kappa} \int_{}^{} \epsilon\, d^3y \sqrt{|\gamma|} (\Kc-\Kc_0) |_{r=r_c}= \]
\[
    =\frac{1}{\kappa} \int_{0}^{\beta}d\tau \int_{S^2}^{} d\Omega^2\, r_c^2 \sqrt{f(r_c)}(\Kc-\Kc_0)|_{r=r_c}= \]
\[
    =\int_{0}^{\beta}d\tau\, \frac{r_c^2}{2}\sqrt{1-\frac{2M}{r_c}}\left[ \frac{-3M+2\left(1-\sqrt{1-\frac{2M}{r}}\right)r}{r^2\sqrt{1-\frac{2M}{r}}}\right]_{r=r_c} = \]
\begin{equation}\label{GHYext}
= \int_{0}^{\beta}d\tau\, \left(\frac{1}{2}\right)\left[ -3M+2\left(1-\sqrt{1-\frac{2M}{r}}\right)r\right]_{r=r_c} = \int_{0}^{\beta}d\tau\, \left(-\frac{M}{2}\right),
\end{equation}
where we have utilised the limit
\begin{equation}
  \lim_{r_c\to\infty} \left(\frac{1}{2}\right)\left(-3M+2\left(1-\sqrt{1-\frac{2M}{r_c}}\right)r_c \right)= -\frac{M}{2} .  
\end{equation}
To fully draw a comparison with the teleparallel approach in the upcoming section, we next ought to evaluate the GHY boundary term residing in the interior of the black hole - also known as the late-time Wheeler-de Witt patch \cite{Krssak:2023nrw}. The important difference now is that we can perform the computation of the action in the Lorenztian signature, as has been recently realised by Susskind et al.\cite{Brown:2015bva, Brown:2015lvg}. Moreover, there is no need for background subtraction, as we are integrating over a compact domain. In turn, there will be two boundaries present: $r_c=0$ and $r_c=2M$. Hence, what we now want to evaluate is the following
\[
    \Ssc_{\text{GHY}}=-\frac{1}{\kappa} \int_{}^{} \epsilon\,d^3y \left[ \sqrt{|\gamma|}  \Kc \right]^{r_c=2M}_{r_c=0}= \frac{1}{\kappa} \int_{}^{}dt \int_{S^2}^{} d\Omega^2 \left[ r_c^2 \sqrt{1-\frac{2M}{r_c}} \Kc  \right]^{2M}_{0}=
    \]
\[
    =\frac{1}{2} \int_{}^{}dt \left[ r_c^2 \left( 1-\frac{2M}{r_c} \right) \left( \frac{M}{r_c^2-2Mr_c}+\frac{2}{r_c}\right)\right]^{2M}_{0}=\int_{}^{}dt \left[r-M \right]^{2M}_{0}=
    \]
\begin{equation}\label{GHYint}
     =\int_{}^{}dt\, 2M .
\end{equation}
which is exactly the result obtained in \cite{Brown:2015lvg}.

\section{Schwarzschild in TEGR}

Before we demonstrate TEGR in practice, let us firstly comment on an apparent circularity in the logic of solving its field equations \cite{Krssak:2015rqa}. The starting point is to specify the spin connection by the use of reference tetrad, which is in turn obtained as a solution to the field equations. However, to write down the actual equations, one also needs to further specify the torsion, which we already know is a function of both the tetrad and the spin connection, hence the circle encloses. To resolve this issue consistently, we adopt an approach not dissimilar to the tetrad formulation of GR: take an ansatz tetrad such that it recovers the ansatz metric, which is postulated from the explicit symmetries of the problem at hand. Such a choice of the tetrad is by no means canonical, as the tangent space Minkowki metric is invariant under local Lorentz transformations. But here the analogy meets its limitations, as in the case of GR, the Levi-Civita spin connection is determined from the tetrad in an unique manner. In TG, we need some other means to further specify the inertial spin connection in order for us to find the torsion, and subsequently the field equations. Recall however that with regards to the equations themselves, we may just as well let the connection vanish. This allows us to postpone the specification of the TG spin connection until we find the tetrad as a solution to the field equations. Only after that we employ the whole machinery of the reference tetrad to obtain the desired associated spin connection.

Let us now move to a particular example of a static spherically symmetric solution along the same lines as in \cite{Krssak:2018ywd} to demonstrate the teleparallel framework in practice. This motivates us to propose the following ansatz for the metric based on the explicit spherical symmetry
\begin{equation}
    ds^2=A^2(r) dt^2 - B^2(r) dr^2 - r^2 d\theta ^2 - r^2 \sin^2 \theta d \phi ^2, 
\end{equation}
with $A(r)$ and $B(r)$ being some unknown functions\footnote{Even though we have already established the Schwarzschild solution previously, for pedagogical reasons, let us proceed with the explicit computation as if the solution is unknown a priori.}, which are to be specified upon solving the field equations. As for the tetrad, most natural choice would be that of the diagonal form
\begin{equation}
    {h^a}_\mu = \text{diag}(A,B,r,r \sin \theta),
\end{equation}
although any other Lorentz transformed frame would make do. To illustrate this fully, let us consider in addition to diagonal tetrad also an off-diagonal one
\begin{equation}
    \Tilde{h}^a{}_\mu = 
        \begin{pmatrix}
        A & 0 & 0 & 0\\
        0 & B \cos\phi \sin \theta & r \cos\phi \cos\theta & -r \sin\phi \sin\theta\\
        0 & -B \cos\theta & r \sin\theta & 0\\
        0 & B \sin\phi \sin\theta & r \sin\phi \cos\theta & r \cos\phi \sin\theta
        \end{pmatrix},
\end{equation}
related to the diagonal tetrad $\Tilde{h}^a{}_\mu \equiv \Tilde{\Lambda}^a{}_b (x) {h^b}_\mu$ by the following local Lorentz transformation 
\begin{equation}
     \Tilde{\Lambda}^a{}_b (x) =
     \begin{pmatrix}
        1 & 0 & 0 & 0\\
        0 & \cos\phi \sin\theta & \cos\phi \cos\theta & -\sin\phi\\
        0 & -\cos\theta & \sin\theta & 0\\
        0 & \sin\phi \sin\theta & \sin\phi \cos\theta & \cos\phi 
        \end{pmatrix}.
\end{equation}

\subsection{Diagonal ansatz}

Next we consider the field equations in the spacetime form accompanied with the assumption of vanishing spin connection, firstly for the case of a diagonal ansatz $h^a{}_\mu$. The non-vanishing components of the superpotential $\Sb_\mu{}^{\rho \sigma}=\Sb_\mu{}^{\rho \sigma}({h^a}_\mu,0)$, antisymmetric in the latter 2 indices, are then
\begin{equation}
    \Sb_t{}^{t r}=-\frac{2}{rB^2}, \quad \Sb_t{}^{t \theta}=\Sb_r{}^{r \theta}=-\frac{\cot \theta}{r^2}, \quad \Sb_\theta{}^{r \theta}=\Sb_\phi{}^{r \phi}=\frac{1}{rB^2}+\frac{A'}{AB^2}.
\end{equation}
Similarly, the non-vanishing components of the energy-momentum pseudotensor $\tb_\mu{}^\rho=\tb_\mu{}^\rho({h^a}_\mu,0)$ are found to be
   \[ \tb_t{}^t=-\tb_r{}^r=\tb_\theta{}^\theta=\tb_\phi{}^\phi=\frac{1}{\kappa}\frac{A+2rA'}{r^2AB^2}, \]
\begin{equation}
    \tb_r{}^\theta=-\frac{1}{\kappa}\frac{BA'+AB'}{r^2AB} \cot \theta, 
\end{equation}
    \[ \tb_\theta{}^r=-\frac{1}{\kappa}\frac{A+rA'}{rAB^2} \cot \theta. \]
Putting it all together, we arrive at the following non-trivial components of the Euler-Lagrange expression
    \[ {E_t}^t = (\frac{-B+B^3+2rB'}{r^2B^3})h, \]
\begin{equation}
    {E_r}^r = (\frac{-A+AB^2-2rA'}{r^2AB^2})h,
\end{equation}
    \[ {E_\theta}^\theta = {E_\phi}^\phi = (\frac{B'(A+rA')-B(A'+rA'')}{rAB^3})h. \]
With an analogy to GR, one verifies that the third equation is indeed a linear combination of the previous ones, so that we are left with a system of two equations with two unknown functions $A=A(r)$ and $B=B(r)$. Namely, let us take a following linear combination
\begin{equation}
    {E_t}^t-{E_r}^r=\frac{2}{r}\frac{(AB)'}{AB^3}=0 \Rightarrow AB=\text{const.}
\end{equation}
from which one can infer, after taking the asymptotic flatness into consideration, the inverse relation $A=\frac{1}{B}$. Plugging this into the third equation, one arrives at
\begin{equation}
    A=\frac{1}{B}=\sqrt{1-\frac{C}{r}},
\end{equation}
with the constant C being specified upon taking the Newtonian limit as
\begin{equation}
    A=\frac{1}{B} \equiv \sqrt{f} \equiv \sqrt{1-\frac{2M}{r}},
\end{equation}
with $M$ representing the mass of the black hole.

Lo and behold, we have just arrived at the well-known and studied Schwarzschild solution \eqref{schwmetric}
\[
    ds^2=f(r) dt^2 - f^{-1}(r) dr^2 - r^2 d\theta ^2 - r^2 \sin^2 \theta d \phi ^2 .
\]

\subsection{Off-diagonal ansatz}
Now let us return to the off-diagonal tetrad $\Tilde{h}^a{}_\mu$, in which case the non-trivial components of superpotential $\Sb_\mu{}^{\rho \sigma}=\Sb_\mu{}^{\rho \sigma}(\Tilde{h}^a{}_\mu,0)$ read 
\begin{equation}
    \Sb_t{}^{t r}=\frac{2(B-1)}{rB^2}, \quad \Sb_\theta{}^{r \theta}=\Sb_\phi{}^{r \phi}=\frac{-A(B-1)+rA'}{rAB^2}.
\end{equation}
Analogously, we find the following components of the energy-momentum pseudotensor $\tb_\mu{}^\rho=\tb_\mu{}^\rho(\Tilde{h}^a{}_\mu,0)$
\[ \tb_t{}^t=\frac{1}{\kappa}\frac{(B-1)(A(B-1)-2rA')}{r^2AB^2}, \quad \tb_r{}^r=\frac{1}{\kappa}\frac{A(B^2-1)-2rA'}{r^2AB^2}, \]
\begin{equation}
    \tb_\theta{}^\theta=\tb_\phi{}^\phi=-\frac{1}{\kappa}\frac{A(B-1)+r(B-2)A'}{r^2AB^2}, \quad \tb_\theta{}^r=\frac{1}{\kappa}\frac{A(B-1)-rA'}{rAB^2} \cot \theta. 
\end{equation}
Comparing these results to the ones obtained for the diagonal tetrad, we see striking differences. No wonder, since by considering the vanishing spin connection in both cases, the mathematical objects of superpotential and pseudotensor  fail to transform covariantly under local Lorentz transformations. Nevertheless, the Euler-Lagrange expression obtained from the off-diagonal ansatz is exactly the same as for the diagonal one, due to the absence of equations governing the dynamics of the spin connection. Moreover, this form is fully equivalent to the Euler-Lagrange expression in general relativity, i.e. the Einstein tensor $\stackrel{\circ}{G}_{\mu \nu}$, which depends solely on the tetrad $h^a{}_\mu$ (or rather on the spacetime metric $g_{\mu \nu}$). 

\subsection{Teleparallel spin connections for a given ansatz}
Here we readily employ our method for finding the spin connection naturally associated to a given tetrad. Firstly, we define a reference tetrad for the diagonal tetrad
\begin{equation}\label{reftetrad}
    {h^a_{(r)}}_\mu \equiv {h^a}_\mu |_{G \rightarrow 0} = \text{diag}(1,1,r,r \sin \theta).
\end{equation}
The knowledge of the reference tetrad then allows for the computation of the coefficients of anholonomy, which in turn define the spin connection. In the diagonal case, we get the following non-vanishing components
\begin{equation}
    \omegab^1{}_{2 \theta}=-1, \quad  \omegab^1{}_{3 \phi}=-\sin\theta, \quad  \omegab^2{}_{3 \phi}=-\cos\theta.
\end{equation}
Along the same lines, we proceed with the construction of the reference tetrad for the off-diagonal ansatz
\begin{equation}
    \Tilde{h}^a_{(r) \mu} \equiv \Tilde{h}^a{}_\mu |_{G \rightarrow 0} = 
        \begin{pmatrix}
        1 & 0 & 0 & 0\\
        0 & \cos\phi \sin\theta & r \cos\phi \cos\theta & -r \sin\phi \sin\theta\\
        0 & -\cos\theta & r \sin\theta & 0\\
        0 & \sin\phi \sin\theta & r \sin\phi \cos\theta & r \cos\phi \sin\theta
        \end{pmatrix}.
\end{equation}
Interestingly enough, this time around the spin connection associated to the off-diagonal tetrad vanishes
\begin{equation}
    \Tilde{\omegab}{}^a{}_{b \mu}=0.
\end{equation}
The form of the local Lorentz transformation relating the diagonal tetrad to non-diagonal one $\Tilde{h}^a{}_\mu = \Tilde{\Lambda}^a{}_b (x) {h^b}_\mu$ ought to be clear now, albeit in a more subtle way. To illuminate this fully, let us consider a reference tetrad in Cartesian coordinates 
\begin{equation}
     h'^a_{(r) \mu} = \text{diag}(1,1,1,1).
\end{equation}
 Next, let us perform a local coordinate transformation from Cartesian to spherical polar coordinate system. This, however, reshuffles the components of the tetrad $h'^a_{(r) \mu}$ to a non-diagonal form $\Tilde{h}^a_{(r) \mu}$, which is in turn brought to the desired form
 \begin{equation}
     h^a_{(r) \mu} = \text{diag}(1,1,r,\sin \theta),
 \end{equation}
 precisely by the inverse of the local Lorentz transformation $\Tilde{\Lambda}^a{}_b$, explaining why the off-diagonal tetrad is indeed the proper one (the spin connection transforms as a proper 1-form w.r.t. coordinate transformations, and as it vanishes in the Cartesian system, the same holds true for the spherical setting), while the diagonal tetrad is stuck with a non-vanishing spin connection due to inhomogeneous transformation law under local Lorentz transformations.

To prove yet another point, let us present the reader with the following expressions for the superpotential $\Sb_\mu{}^{\rho \sigma}({h^a}_\mu,\omegab^a{}_{b \mu})$ and the energy-momentum pseudotensor $\tb_\mu{}^\rho({h^a}_\mu,\omegab^a{}_{b \mu})$, now functions of both the diagonal tetrad and the appropriate spin connection, whose components finally match with those obtained in off-diagonal case, reflecting the tensorial nature of the quantities in question
\begin{equation}
    \Sb_t{}^{t r}=\frac{2(B-1)}{rB^2}, \quad \Sb_\theta{}^{r \theta}=\Sb_\phi{}^{r \phi}=\frac{-A(B-1)+rA'}{rAB^2},
\end{equation}
\[ \tb_t{}^t=\frac{1}{\kappa}\frac{(B-1)(A(B-1)-2rA')}{r^2AB^2}, \quad \tb_r{}^r=\frac{1}{\kappa}\frac{A(B^2-1)-2rA'}{r^2AB^2}, \]
\begin{equation}
    \tb_\theta{}^\theta=\tb_\phi{}^\phi=-\frac{1}{\kappa}\frac{A(B-1)+r(B-2)A'}{r^2AB^2}, \quad \tb_\theta{}^r=\frac{1}{\kappa}\frac{A(B-1)-rA'}{rAB^2} \cot \theta. 
\end{equation}
 \subsection{Regularisation of the action}
Now we are ready to fully demonstrate the benefits of considering the spin connection when evaluating the action. But firstly, let us briefly stop by the case, when we unwittingly assume a vanishing connection for the diagonal tetrad
\begin{equation}
    {h^a}_\mu = \text{diag}(f^{\frac{1}{2}},f^{-\frac{1}{2}},r,r \sin \theta), \quad f=f(r)=1-\frac{2M}{r}.
\end{equation}
This action, as we are about to see, turns out divergent precisely due to us neglecting the inertial effects present in such a frame
\begin{equation}
    \Ssb (h^a{}_\mu,0)=\frac{1}{\kappa}\int_{\mathcal{M}}^{} d^4x \sin \theta \longrightarrow \infty.
\end{equation}
On the other hand, if we remove those pesky inertial effects by considering the appropriate  spin connection (or we aptly work with a proper frame from the beginning), we arrive at what can be viewed as a renormalised action. Teleparallel approach thus predicts the value of the action in the Euclidean patch (exterior region $r\in [2M,\infty)$) as
   \[ \Ssb_{\text{ren}}(h^a{}_\mu, \omegab^a{}_{b \mu})=\Ssb_{\text{ren}}(\Tilde{h}{}^a{}_\mu, 0)=\frac{1}{\kappa}\int_{\text{ext}}^{} d^4x (f^{\frac{1}{2}}-1)(f^{\frac{1}{2}}-1+2r(f')^{\frac{1}{2}}) \sin \theta = \]
    \[ =\frac{1}{\kappa}\int_{}^{} dt \int_{2M}^{\infty} dr \int_{S^2}^{} d\Omega \frac{(-1+\sqrt{1-\frac{2M}{r}})(2M+(-1+\sqrt{1-\frac{2M}{r}})r)}{r-2M} = \]
\begin{equation}
    = \int_{}^{} dt \left( -M \right),
\end{equation}
which is, however, twice the result obtained in \eqref{GHYext}.

Similarly, we continue to evaluate the action on the late-time Wheeler-de Witt patch (interior region $r\in [0,2M]$) as
\[ \Ssb_{\text{ren}}(h^a{}_\mu, \omegab^a{}_{b \mu})=\Ssb_{\text{ren}}(\Tilde{h}{}^a{}_\mu, 0)=\frac{1}{\kappa}\int_{\text{int}}^{} d^4x (f^{\frac{1}{2}}-1)(f^{\frac{1}{2}}-1+2r(f')^{\frac{1}{2}}) \sin \theta = \]
    \[ =\frac{1}{\kappa}\int_{}^{} dt \int_{0}^{2M} dr \int_{S^2}^{} d\Omega \frac{(-1+\sqrt{1-\frac{2M}{r}})(2M+(-1+\sqrt{1-\frac{2M}{r}})r)}{r-2M} = \]
\begin{equation}
    = \int_{}^{} dt\, 2M,
\end{equation}
we see that this time around, we obtain a matching result as in \eqref{GHYint}.

\begin{figure}[!ht]
\centering
\includegraphics[width=.6\linewidth]{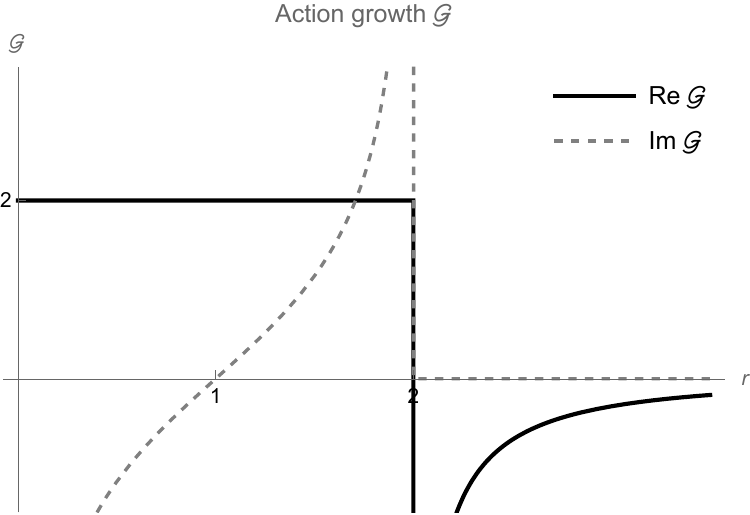}
\captionof{figure}{Graph of the rescaled teleparallel action growth $\mathcal{G}=\kappa \frac{d \mathcal{S}}{dt}$ for the Schwarzschild solution \cite{Krssak:2023nrw}. Real part of the function is denoted with the black full line, while the dotted gray line represents the imaginary part.}
\end{figure}

Since here we are interested in a volume integral, this then allows us to graph the teleparallel Lagrangian to further investigate the behaviour of the subintegral function in the respective regions of the black hole as demonstrated above.

\section{Canonical frame \label{Canframe}}

 As has been shown already, the GHY boundary term in Riemannian geometry leads to the value of $2M$ in the interior, and $-\frac{M}{2}$ in the exterior region of the Schwarzschild black hole. Interestingly, we have obtained a matching result of $2M$ for the interior region within the teleparallel geometry. Nonetheless, in the exterior region, these two approaches seem to differ by a factor of 2, as teleparallel gravity predicts the action to be proportional to the value of $-M$. In this section, we shall thus investigate another class of frames to possibly discern between the two contrasting approaches of teleparallel gravity and general relativity. 

\subsection{Kerr-Schild ansatz}

 To this end, let us now examine the claim made by Koivisto et.al.\cite{BeltranJimenez:2019bnx}, who argue for the use of the so-called canonical frame when it comes to the evaluation of the black hole action. This particular frame is defined by a rather simple condition of vanishing energy-momentum pseudotensor
 \begin{equation}
     t^\mu{}_\nu = 0.
 \end{equation}
 Luckily for us, there is a special class of frames that solves the above condition. These are included in the Kerr-Schild type of metrics of a following form \cite{Frob:2021dmv}
\begin{equation}
    g_{\mu \nu} = \bar{g}_{\mu \nu} - 2Fk_\mu k_\nu, \quad F=\frac{Mr^3}{r^4+a^2 z^2},
\end{equation}
where $\bar{g}_{\mu \nu}$ stands for the background flat Minkowski metric in Cartesian coordinates. Null-vector $k^\mu$ is defined as $g^{\mu \nu}k_\mu k_\nu \equiv \bar{g}^{\mu \nu}k_\mu k_\nu =0$. Historically, this solution lead to the description of a rotating black hole by Roy Kerr, who utilised the oblate spheroidal coordinates. In this better suited system, the deviation from spherical form is encoded in the parameter $a$
\[
    x=\sqrt{r^2+a^2} \sin{\theta} \cos{\phi}, \quad y=\sqrt{r^2+a^2} \sin{\theta} \sin{\phi}, \quad z=r \cos{\theta}.
\]
Nevertheless, for our purposes, it will be sufficient to work within the non-rotating limit characterised by vanishing parameter $a=0$. Consequently, the background flat metric takes on a spherical form
\[
    \bar{g}_{\mu \nu}=\text{diag}(1,-1,-r^2,-r^2\sin^2{\theta}).
\]
Null (co-)vector $k_\mu$ and function F are found to be
\[
    k_\mu=(1,1,0,0), \quad F=\frac{M}{r},
\]
which leads us ultimately to the Schwarzschild solution in Eddington-Finkelstein (EF) coordinates
\begin{equation}\label{EFmetric}
    g_{\mu \nu}=\bar{g}_{\mu \nu} -2F k_\mu k_\nu =  
    \begin{pmatrix}
        1-\frac{2M}{r}& -\frac{2M}{r} & 0 & 0\\
        -\frac{2M}{r} & -1-\frac{2M}{r} & 0 & 0\\
        0 & 0 & -r^2 & 0\\
        0 & 0 & 0 & -r^2 \sin^2 \theta
    \end{pmatrix}.
\end{equation}

\subsection{Canonical frame in TEGR}

Let us now switch to the teleparallel geometry and establish the fundamental variables: tetrad and spin connection. From the metric decomposition into background flat metric and a part quadratic in  null-vector, we get analogous expression for the full tetrad $h^a{}_\mu$
\begin{equation}
    g_{\mu \nu}=\bar{g}_{\mu \nu} -2Fk_\mu k_\nu = \eta_{a b} h^a{}_\mu h^b{}_\nu \Rightarrow h^a{}_\mu=\bar{h}^a{}_\mu -Fk_\mu k^a,
\end{equation}
where we have defined the background tetrad $\bar{h}^a{}_\mu$, which in the non-rotating limit relates the spacetime Minkowski metric in the spherical coordinates to the tangent space Minkowski metric in the Cartesian system
\begin{equation}
    \bar{h}^a{}_\mu=\text{diag}(1,1,r,r \sin \theta).
\end{equation}
Additionally, $k^a$ now represents the null-vector in the frame basis
\begin{equation}
    k^a \equiv h^a{}_\mu k^\mu = \bar{h}^a{}_\mu k^\mu = (1,-1,0,0)^T,
\end{equation}
which ultimately leads us to the expression for the full-tetrad, which we shall refer back to as the canonical frame
\begin{equation}
    h^a{}_\mu = 
    \begin{pmatrix}
        1-\frac{M}{r} & -\frac{M}{r} & 0 & 0\\
        \frac{M}{r} & 1+\frac{M}{r} & 0 & 0\\
        0 & 0 & r & 0\\
        0 & 0 & 0 & r \sin \theta
    \end{pmatrix}.
\end{equation}
To specify the associated spin connection, all we need is to realise that it is determined by the background flat tetrad $\bar{h}^a{}_\mu$, which we have already dealt with in preceding sections \eqref{reftetrad}; or equivalently, employ the procedure of reference tetrad. Either way, the only non-vanishing components of $\bar{\omega}^a{}_{b \mu}$ turn out to be
\begin{equation}\label{properspincon}
    \bar{\omega}^1{}_{2 \theta}=-1, \quad  \bar{\omega}^1{}_{3 \phi}=-\sin\theta, \quad  \bar{\omega}^2{}_{3 \phi}=-\cos\theta,
\end{equation}
which demonstrates the fact that we have moved from the inertial Cartesian frame to a non-inertial spherical one. Before we move to calculate the torsion and evaluate the action for this canonical frame, we may still perform one final coordinate transformation from EF coordinate system to Schwarzschild one
\begin{equation}
    t \rightarrow t + 2M \text{ln} (r-2M) \Longrightarrow dt \rightarrow dt + \frac{2M}{r-2M}dr,
\end{equation}
which brings the metric to the already familiar diagonal form 
\begin{equation}
    g_{\mu \nu}= 
     \begin{pmatrix}
        \left(1-\frac{2M}{r}\right) & 0 & 0 & 0\\
        0 & -\left(1-\frac{2M}{r}\right)^{-1} & 0 & 0\\
        0 & 0 & -r^2 & 0\\
        0 & 0 & 0 & -r^2 \sin^2 \theta
    \end{pmatrix}.
\end{equation}
Such a transformation effects only the 0 and 1 component of the canonical tetrad
\begin{equation}
    h'^a{}_\mu=
    \begin{pmatrix}
        1-\frac{M}{r} & \frac{M}{r-2M} & 0 & 0\\
        \frac{M}{r} & \frac{r-M}{r-2M} & 0 & 0\\
        0 & 0 & r & 0\\
        0 & 0 & 0 & r \sin \theta
    \end{pmatrix}.
\end{equation}
To fully draw a comparison with our previous results, we can bring this tetrad to a diagonal form $\Tilde{h}^a=\Lambda^a{}_b h'^b$ by the following Lorentz transformation
\begin{equation}
    \Lambda^a{}_b=
    \begin{pmatrix}
        \frac{r-M}{\sqrt{r(r-2M)}} & -\frac{M}{\sqrt{r(r-2M)}} & 0 & 0\\
         -\frac{M}{\sqrt{r(r-2M)}} & \frac{r-M}{\sqrt{r(r-2M)}} & 0 & 0\\
        0 & 0 & 1 & 0\\
        0 & 0 & 0 & 1
    \end{pmatrix},
\end{equation}
which yields precisely the diagonal tetrad we have used in previous sections
\begin{equation}
    \Tilde{h}^a{}_\mu=\text{diag} \left( \sqrt{1-\frac{2M}{r}},\frac{1}{\sqrt{1-\frac{2M}{r}}},r,r \sin \theta \right).
\end{equation}
The spin connection is, needles to say, not affected be any of these transformations. Although the coordinate shift from E-F to Schwarzschild coordinates reshuffled the components of the tetrad, the limit $r \rightarrow \infty$ ensures that the reference tetrad is in both cases identical. Moreover, the spin connection could acquire an inhomogeneous term after performing the LLT, but within the same limit as before, the corresponding transformation reduces to identity, and the spin connection is ultimately left unchanged. We thus see a first hint at the ambiguity in assigning a spin connection to a given tetrad, as $\bar{\omega}^a{}_{b \mu}$ corresponds to the canonical frame and the diagonal tetrad at the same time. 

\subsection{Canonical frame and the role of Stoke's theorem \label{chapcanframe}}

To illuminate this ambiguity of spin connection, let us move to the evaluation of action for the canonical frame $h^a{}_\mu$. Firstly, the non-vanishing components of torsion $T^a{}_{\mu \nu}(h^a{}_\mu,\bar{\omega}^a{}_{b \mu})$ are
\begin{equation}
    T^0{}_{t r}=-T^1{}_{t r}=-\frac{M}{r^2}, \quad T^2{}_{t \theta}=T^2{}_{r \theta}=-\frac{M}{r}, \quad T^3{}_{t \phi}=T^3{}_{r \phi}=-\frac{M \sin \theta}{r}.
\end{equation}
Interestingly, despite the torsion tensor being non-vanishing, the torsion scalar constructed from it does indeed vanish
\begin{equation}
    T=\frac{1}{4} T^\rho{}_{\mu \nu} T_\rho{}^{\mu \nu}+\frac{1}{2} T^\rho{}_{\mu \nu} T^{\nu \mu}{}_\rho - T^\rho{}_{\mu \rho} T^{\nu \mu}{}_\nu = 0.
\end{equation}
The action is consequently trivially identical to zero 
\begin{equation}
    \Ssb(h^a{}_\mu,\bar{\omega}^a{}_{b \mu})=\int^{}_{} \frac{h}{2 \kappa} T = 0 .
\end{equation}
Hence the pair of canonical variables $\{ h^a{}_\mu,\bar{\omega}^a{}_{b \mu} \}$ does fulfill the role of regularising the BH action, albeit it does so in a maximal fashion, rendering the action impractical in the process. To fully draw a comparison with the procedure established in \cite{BeltranJimenez:2019bnx}, let us recall a decomposition of teleparallel Lagrangian into the EH Lagrangian of general relativity with respect to the Riemannian connection and a total divergence term
\begin{equation}
    \mathcal{L}_{TG}=\Lc_{EH}-\partial_\mu \left(\frac{h}{\kappa}T^\mu\right).
\end{equation}
Realising that both the Ricci scalar for the Schwarzschild solution and the torsion scalar corresponding to the canonical frame vanish, the above equality trivially resolves for a vanishing total divergence. Indeed, the determinant of the canonical frame $h \equiv \text{det}(h^a{}_\mu)$ and the non-vanishing components of the vector torsion $T^\mu \equiv T^{\nu \mu}{}_\nu$ are found to be
\begin{equation}\label{cfvector}
    h = r^2 \sin{\theta}, \quad T^r=-T^t=\frac{M}{r^2},
\end{equation}
resulting in a vanishing total divergence term. How is then possible to obtain a non-zero prediction, if the canonical frame indubitably leads to a trivial black hole action? To fully illuminate this conundrum, let us transform the total divergence to a boundary term by utilising the Stoke's theorem, which mirrors the approach of GHY boundary terms within general relativity \cite{Oshita:2017nhn,BeltranJimenez:2019bnx}. Just as before, it is then customary to focus on the Euclidean version of the Schwarzschild action by the means of Wick's rotation $t \rightarrow \tau=it$, which compactifies the time coordinate on the interval $\tau \in [0,\beta=8\pi M]$, removing the conical singularity at the event horizon in the process. Once again, we are interested in the sole horizon defined by the hypersurface $r=r_c$, with the constant $r_c$ sent to infinity. What we would like to evaluate then is along the following lines
\begin{equation}
    \Ssb_{\text{bound}}=-\frac{1}{\kappa}\int_{r_c \rightarrow \infty}^{} d^3y \epsilon  \sqrt{|\gamma|} n_\mu  (T^\mu - T^\mu_0) |_{r=r_c} ,
\end{equation}
where $n_\mu=-\delta_{\mu r}\left(1-\frac{2M}{r}\right)^{-\frac{1}{2}}$ is a unit normal to the aforementioned hypersurface. This form of unit normal is by no means surprising, as it coincides with the expression found following the GHY approach due to the hypersurface being set up in the same way in both cases. At the same time, the metric fails to remain diagonal in the EF coordinate system, and therefore we expect the normal vector to acquire non-vanishing component in the temporal direction when expressed in the cotravariant form. Indeed, if we raise the index with the inverse of the metric \eqref{EFmetric}
\begin{equation}\label{invEFmetric}
    g^{\mu \nu}=\bar{g}^{\mu \nu} +2F k^\mu k^\nu =  
    \begin{pmatrix}
        1+\frac{2M}{r}& -\frac{2M}{r} & 0 & 0\\
        -\frac{2M}{r} & -1+\frac{2M}{r} & 0 & 0\\
        0 & 0 & -\frac{1}{r^2} & 0\\
        0 & 0 & 0 & -\frac{\csc^2{\theta}}{r^2}
    \end{pmatrix}.
\end{equation}
we find that 
\begin{equation}\label{normalvectorEF}
    n^\mu=g^{\mu \nu} n_\nu=\sqrt{f}(f^{-1}-1,1).
\end{equation}

Now we are ready to specify the induced metric $\gamma_{i j}= \gamma_{\mu \nu} e^\mu{}_i e^\nu{}_j$ on the said hypersurface. The coordinates on this boundary are yet again chosen as $y^i=(t,\theta, \phi)$, which results in the following projectors $e^\mu{}_i$ in a 4x3 matrix form
\begin{equation}
    e^\mu{}_i = 
    \begin{pmatrix}
        1 & 0 & 0 \\
        0 & 0 & 0 \\
        0 & 1 & 0 \\
        0 & 0 & 1
    \end{pmatrix},
\end{equation}
leading us to the desired form of the induced 3-metric on the hypersurface $r=r_c$
\begin{equation}
    \gamma_{i j }=
    \begin{pmatrix}
        1-\frac{2M}{r} & 0 & 0\\
        0 & -r^2 & 0\\
        0 & 0 & -r^2 \sin^2 \theta
    \end{pmatrix}, \quad \sqrt{|\gamma|}= \sqrt{f(r_c)}\,r_c^2 \sin{\theta}.
\end{equation}
Returning back to the expression of boundary action, the final observation is in place for the counter term $T^\mu_0$ for the asymptotically flat background. It turns out that due to us assuming the spin connection $\bar{\omega}^a{}_{b \mu}$ constructed from the reference tetrad, this regulating term trivially vanishes, highlighting the role of such connection in action regularisation. With all the important ingredients in place, we move to the evaluation of action
\[
    \Ssb_{\text{bound}}=-\frac{1}{\kappa}\int_{}^{} d^3y \epsilon  \sqrt{|\gamma|} n_r  T^r |_{r=r_c}=
\]
\[
 =-\frac{1}{\kappa} \int_{0}^{\beta}d\tau\, \int_{S^2}^{} d\Omega^2\, \epsilon r_c^2 \sqrt{1-\frac{2M}{r_c}} n_r  T^r= 
\]
\begin{equation}
   =-\frac{1}{2} \int_{0}^{\beta}d\tau\, r_c^2  \left(\frac{M}{r_c^2}\right)= \int_{0}^{\beta}d\tau\, \left( -\frac{M}{2} \right),
\end{equation}
which recovers the prediction for the exterior region of the black hole following the GHY approach. The interior region, on the other hand, remains trivial
\[ \Ssb_{\text{bound}}=-\frac{1}{\kappa}\int_{}^{} d^3y \epsilon  \left[ \sqrt{|\gamma|} n_\mu  T^\mu \right]^{r_c=2M}_{r_c=0}=\]
\begin{equation}\label{cfintsol}
=\int_{}^{}dt\, \left[ -\frac{M}{2} \right]^{r_c=2M}_{r_c=0}=0 .
\end{equation}

\subsection{GHY term for Schwarzschild in E-F coordinates}

For completeness, we present the computation of the GHY boundary term for the Schwarzschild solution in EF coordinates
\begin{equation}\label{efmet}
    g_{\mu \nu}=  
    \begin{pmatrix}
        1-\frac{2M}{r}& -\frac{2M}{r} & 0 & 0\\
        -\frac{2M}{r} & -1-\frac{2M}{r} & 0 & 0\\
        0 & 0 & -r^2 & 0\\
        0 & 0 & 0 & -r^2 \sin^2 \theta
    \end{pmatrix}.
\end{equation}
Fortunately, we are already equipped with most of the results needed for this purpose. Indeed, the unit normal to the hypersurface $r=r_c$ was found in eq.\eqref{normalvectorEF} as
\begin{equation}
    n^\mu=f^{\frac{1}{2}}(f^{-1}-1,1).
\end{equation}
This allows for further specification of the trace of extrinsic curvature $\Kc$
\[  \Kc = n^\mu_{;\mu}=\frac{1}{\sqrt{-g}}\partial_\mu(\sqrt{-g}n^\mu)= \]
  \[ =\frac{1}{r^2 \sin{\theta}} \left[ \partial_t \left( f^{\frac{1}{2}}(f^{-1}-1)r^2 \sin{\theta} \right)+\partial_r \left(\ f^{\frac{1}{2}}r^2 \sin{\theta}\right)\right]= \]
 \[
  = \sqrt{1-\frac{2M}{r}} \left( \frac{M}{r^2-2Mr}+\frac{2}{r}\right)
  \]
\begin{equation}
    = \frac{1}{r^2\sqrt{f}} \left( -3M+2r \right),
\end{equation}
which is precisely the expression we got in eq. \eqref{extcurvschw} due to $t$ being a cyclic coordinate. In other words, despite the normal vector acquiring a time component in contrast to the case of Schwarzschild solution in spherical coordinates, derivative with respect to $t$ is trivial due to the absence of explicit dependency of the component on temporal coordinate.

Along the same lines as in Section ~\ref{chapcanframe}, we find the 3-metric induced on the boundary $r=r_c$ as
\begin{equation}
    \gamma_{i j }=
    \begin{pmatrix}
        1-\frac{2M}{r} & 0 & 0\\
        0 & -r^2 & 0\\
        0 & 0 & -r^2 \sin^2 \theta
    \end{pmatrix}, \quad \sqrt{|\gamma|}= \sqrt{f(r_c)}\, r_c^2 \sin{\theta}.
\end{equation}
What requires a further comment, however, is that now we need to regularise the action by the background subtraction in contrary to the method of boundary term within teleparallel gravity, where the same end was reached by the means of specification of the spin connection associated to a given reference tetrad.

To this end, we utilise the limit $M \rightarrow 0$ in the $g_{tt}$, $g_{rr}$, and also the off-diagonal components $g_{t r}$ of the metric \eqref{efmet}, which results in the Minkowski metric in the spherical coordinates
\begin{equation}
    \eta_{\mu \nu} =  
    \begin{pmatrix}
        1& 0 & 0 & 0\\
        0 & -1 & 0 & 0\\
        0 & 0 & -r^2 & 0\\
        0 & 0 & 0 & -r^2 \sin^2 \theta
    \end{pmatrix},
\end{equation}
leading to the identical expression for the trace of the extrinsic curvate for the asymptotically flat background $\Kc_0$ as in eq. \eqref{extcurvflatbcg}
\begin{equation}
    \Kc_0=\frac{2}{r}.
\end{equation}
Let us recall that an analogous situation has occurred also in the case of canonical frame way back in \eqref{properspincon}, where we have identified the same spin connection to both canonical and the diagonal frame of the Schwarzschild solution following the method of reference tetrad.

The evaluation of action $\Ssc_{\text{GHY,0}}$ for the Schwarzschild solution in EF coordinates thus completely reduces to the same computation as in eq. \eqref{GHYext} and eq. \eqref{GHYint}, predicting the external solution to be proportional to $-\frac{M}{2}$, while the internal solution yet again recovers the value of $2M$.

\section{Boundary term for the diagonal Schwarzschild tetrad}

Surprisingly, following the steps in \cite{BeltranJimenez:2018vdo}, the same magic trick seems to be at play already at the level of the diagonal frame for the Schwarzschild solution
\begin{equation}
     \Tilde{h}^a{}_\mu = \text{diag}\left( \sqrt{f},\frac{1}{\sqrt{f}},r,r \sin \theta \right),
\end{equation}
while considering the following spin connection $\Tilde{\omega}^a{}_{b \mu}$ obtained by the method of reference tetrad
\begin{equation}
    \Tilde{\omega}^1{}_{2 \theta}=-1, \quad  \Tilde{\omega}^1{}_{3 \phi}=-\sin\theta, \quad  \omega^2{}_{3 \phi}=-\cos\theta.
\end{equation}
Within this particular diagonal frame, the only non-vanishing component of the vector torsion is found to be
\begin{equation}\label{vectorschw}
    T^r=\frac{3M+2\left(-1+\sqrt{1-\frac{2M}{r}}\right)r}{r^2}.
\end{equation}
Yet again, we compute the action corresponding to the boundary term
\[
    \Ssb_{\text{bound}}=-\frac{1}{\kappa}\int_{}^{} d^3y \epsilon  \sqrt{|\gamma|} n_\mu  T^\mu |_{r=r_c}=
\]
\[
 =-\frac{1}{\kappa} \int_{0}^{\beta} \epsilon\, d\tau \int_{S^2}^{} d\Omega^2\, r_c^2 \sqrt{1-\frac{2M}{r_c}} n_r  T^r= 
\]
\begin{equation}
   =-\frac{1}{2} \int_{0}^{\beta}d\tau\, r_c^2  T^r=\int_{0}^{\beta}d\tau\, \left( -\frac{M}{2} \right),
\end{equation}
where we have utilised the limit $r_c \rightarrow \infty$ for the component of vector torsion
\begin{equation}
 \lim_{r_c\to\infty} \left( -\frac{r^2_c}{2}T^r \right)=\lim_{r_c\to\infty} \left[ \left(-\frac{r^2_c}{2}\right) \cdot \frac{3M+2\left(-1+\sqrt{1-\frac{2M}{r_c}}\right)r_c}{r_c^2} \right]= -\frac{M}{2},   
\end{equation}
which agrees with the prediction for the external region of the black hole according to the procedure of GHY boundary term. Lastly, we check for the internal solution 
\[ \Ssb_{\text{bound}}=-\frac{1}{\kappa}\int_{}^{} d^3y \epsilon  \left[ \sqrt{|\gamma|} n_\mu  T^\mu \right]^{r_c=2M}_{r_c=0}=\]
\[
=\int_{}^{}dt \left[ \left( -\frac{r^2}{2} \right) \frac{3M+2\left(-1+\sqrt{1-\frac{2M}{r}}\right)r}{r^2}\right]^{r_c=2M}_{r_c=0}= \]
\begin{equation}
=\int_{}^{}dt \left[\frac{M}{2}-\left(-\frac{3M}{2}\right)\right]=\int_{}^{}dt\, 2M,
\end{equation}
with the respective limits found as
\begin{equation}
 \lim_{r_c\to 2M} \left[ \left(-\frac{r^2_c}{2}\right) \cdot \frac{3M+2\left(-1+\sqrt{1-\frac{2M}{r_c}}\right)r_c}{r_c^2} \right]= \frac{M}{2},   
\end{equation}
\begin{equation}
 \lim_{r_c\to 0} \left[ \left(-\frac{r^2_c}{2}\right) \cdot \frac{3M+2\left(-1+\sqrt{1-\frac{2M}{r_c}}\right)r_c}{r_c^2} \right]= -\frac{3M}{2}.   
\end{equation}
We thus truly see that we recover the prediction for gravitational action of a Schwarzschild black hole in each region respectively by utilising the decomposition of teleparallel lagrangian into the EH Lagrangian and the total divergence term, but only thanks to the application of the Stokes' theorem.

\chapter{Discussion}
In this chapter, we shall summarise the results for the black hole action according to the respective approaches. Then we shall discuss the main suspect responsible for the apparent inconsistencies within the considered procedures, and conclude with the practical modifications to the models at hand.

We have so-far considered altogether three distinct approaches with a focus on two instances of the Schwarzschild solution: the GH method of general relativity with the metric expressed in Schwarzschild and EF coordinates; and then the teleparallel approach in the sense of both volume and surface integrals within the proper and canonical frame respectively. The results may thus be summarised as follows\footnote{A similar situation, albeit in a slightly different context of conserved quantities, is also present in \cite{Emtsova:2021ehh}. Namely, the Noether charges acquire contrasting values in the Schwarzschild static and Lemaitre gauges based on the various choices of the time-like Killing vector.}

\begin{table}[h!]
\centering
\begin{tabular}{ |c||c|c|c|c|c|c|  }
 \hline
  & \multicolumn{2}{c}{$\Ssc_{\textbf{GHY,0}}$} & \multicolumn{2}{|c}{\textbf{Proper frame}} & \multicolumn{2}{|c|}{\textbf{Canonical frame}} \\
 \hline
 \textbf{BH region/Model} & Schw. coord. & EF coord. & $\Ssb_{\text{TG}}$ & $\Ssb_{\text{bound}}$ & $\Ssb_{\text{TG}}$ & $\Ssb_{\text{bound}}$ \\
 \hline\hline
 \textbf{Interior} & 2M & 2M & 2M & 2M & 0 & 0 \\
 \hline
 \textbf{Exterior} & -M/2 & -M/2 & -M & -M/2 & 0 & -M/2 \\
 \hline
\end{tabular}
\caption{Predictions for the gravitational action with respect to the considered models and within respective regions of the black hole. The particular values are understood in the sense of the action growth $\frac{d \mathcal{S}}{dt}$.}
\label{table:1}
\end{table}

The natural conclusion drawn from these observations could be that the sole perpetrator responsible for the discrepancies within respective theories of teleparallel gravity and general relativity was, afterall, hidden right under our noses: the all-familiar Stokes' theorem. The evidence points way back to the derivation of the GHY boundary term, where the use of the said theorem in variational procedure of EH action was justified on the grounds of variations of the Christoffel symbols $\delta \Gammac^\mu{}_{\rho \sigma}$ being tensorial in nature. Nevertheless, this argument fails in the "practical" setting, outside of the variations of the action. To illuminate this fully, let us split the EH Lagrangian 
\begin{equation}
    \Lc_{\text{EH}}=\frac{1}{2\kappa} \sqrt{-g}\Rc=\Lc_{\text{bulk}}+\Lc_{\text{tot}},
\end{equation}
into a bulk $\Lc_{\text{bulk}}$ and a total derivative term $\Lc_{\text{tot}}$ accordingly
\[
    \Lc_{\text{bulk}}=\frac{1}{2\kappa}\sqrt{-g} g^{\mu \nu} \left( \Gammac^\rho{}_{\sigma \mu}\Gammac^\sigma{}_{\rho \nu}-\Gammac^\rho{}_{\mu \nu}\Gammac^\sigma{}_{\rho \sigma}\right),
\]
\begin{equation}
    \Lc_{\text{tot}}=\frac{1}{2\kappa}\partial_\mu \left[ \sqrt{-g}\left( g^{\nu \sigma}\Gammac^\mu{}_{\nu \sigma}- g^{\mu \nu}\Gammac^\sigma{}_{\nu \sigma}\right)\right]=\frac{1}{2\kappa}\partial_\mu \left(\sqrt{-g} V^\mu \right),
\end{equation}
where we have - by the abuse of notation - denoted the quantity $V^\mu$ in analogy to the variational procedure of EH action \cite{Padmanabhan:2010zzb}
\begin{equation}
    V^\mu=\left( g^{\rho \sigma}\Gammac^\mu{}_{\rho \sigma}- g^{\mu \nu}\Gammac^\sigma{}_{\nu \sigma}\right) \equiv -\frac{1}{g} \partial_\nu (g g^{\mu \nu}).
\end{equation}
The crucial distinction is that this time around, $V^\mu$ does not represent a vector, which has profound consequences in the integration of the action constructed from this total derivative term
\begin{equation}
    \Ssc_{\text{tot}}= \int_{\mathcal{M}}^{} d^4x \Lc_{\text{tot}} = \frac{1}{2\kappa} \int_{\mathcal{M}}^{} d^4x \partial_\mu\left( \sqrt{-g}V^\mu \right).
\end{equation}
Indeed, if we naively transform this term to a boundary term with the help of Stokes' theorem
\begin{equation}
    \Ssc_{\text{bound}}= \frac{1}{2\kappa} \int_{\mathcal{\partial M}}^{} \epsilon d^3y \sqrt{|\gamma|} n_\mu V^\mu,
\end{equation}
we shall quickly come to a realisation that these two terms are not equal
\begin{equation}
    \Ssc_{\text{tot}} \neq \Ssc_{\text{bound}}.
\end{equation}
To this end, let us recall the Schwarzschild metric in the spherical coordinates from eq. \eqref{schwmetric}. The "pseudo"-vector is then found as
\begin{equation}\label{pseudovector}
    V^\mu=\left( 0, \frac{-6M+4r}{r^2},\frac{2 \cot{\theta}}{r^2},0\right).
\end{equation}
The bulk $\Lc_{\text{bulk}}$ and total derivative Lagrangians $\Lc_{\text{tot}}$ are specified followingly
\begin{equation}
    \Lc_{\text{bulk}}=-\frac{1}{\kappa} \sin{\theta}, \quad \Lc_{\text{tot}}=\frac{1}{\kappa} \sin{\theta},
\end{equation}
which correctly results in a vanishing EH action 
\begin{equation}
    \Lc_{\text{EH}}=\Lc_{\text{bulk}}+\Lc_{\text{tot}}=0,
\end{equation}
due to the Schwarzschild solution being a vacuum solution.

On the other hand, carrying out the computation for the $\Ssc_{\text{bound}}$ term by setting the hypersurface as $r=r_c$ with $r_c \rightarrow \infty$, and utilising the results obtained in Section ~\ref{chapghy}
\begin{equation}
    n_\mu =  - \frac{\delta_{\mu r}}{\sqrt{f}},
\end{equation}
and
\begin{equation}
    \sqrt{|\gamma|}=\sqrt{f(r_c)} r^2 \sin{\theta},
\end{equation}
we arrive at the following expression
\[
    \Ssc_{\text{bound}}= \frac{1}{2\kappa} \int_{\mathcal{\partial M}}^{} \epsilon d^3y \sqrt{|\gamma|} n_\mu V^\mu=
\]
\[
    = -\frac{1}{2 \kappa} \int_{\mathcal{\partial M}}^{} dt\, \sqrt{f(r_c)} r_c^2 \sin{\theta} \left( - \frac{1}{\sqrt{f(r_c)}}\right) \left( \frac{-6M+4r_c}{r_c^2}\right)=
\]
\begin{equation}\label{boundterm}
    =\frac{1}{2} \int_{\mathcal{}}^{}  dt\, ( -3M+2r_c),
\end{equation}
which is precisely the GHY boundary term obtained in \eqref{extcurvschw}, without the counter term for the asymptotically flat background. Integrating the bulk term we get
\begin{equation}\label{bulkterm}
    \Ssc_{\text{bulk}}=\int_{\mathcal{M}}^{} \Lc_{\text{bulk}}=-\frac{1}{\kappa} \int_{\mathcal{M}}^{} \sin{\theta}=-\frac{1}{2} \int_{\mathcal{}}^{} dt\,r_c.
\end{equation}
We immediately see that the bulk \eqref{bulkterm} and the boundary term \eqref{boundterm}, do not add to zero
\begin{equation}
    \Ssc_{\text{bulk}}+\Ssc_{\text{bound}}=\frac{1}{2} \int_{\mathcal{}}^{} dt{} (-3M+r_c),
\end{equation}
but rather, this expression diverges in the limit when boundary goes to infinity. Although such a result was expected due to us not considering any regularising methods, the main point now is that we got two distinct solutions based on the employment of Stokes' theorem
\[
    \Ssc_{\text{tot}} \neq \Ssc_{\text{bound}},
\]
\begin{equation}
    \frac{1}{2} \int_{\mathcal{}}^{} dt\, r_c \neq \frac{1}{2} \int_{\mathcal{}}^{}  dt\, ( -3M+2r_c)
\end{equation}
The reason for the contrasting results is our failure in satisfying the assumptions for the valid use of the said theorem, which holds if and only if the the object in question is a regular smooth vector field. In our case, the field $V^\mu$ in eq. \eqref{pseudovector} is, first and foremost, not a vector field as it fails to transform covariantly under diffeomorphisms. Secondly, it is not a regular field on the whole of $\mathbb{R}^{1,3}$, but rather on $\mathbb{R} ^{1,3}\setminus \{ 0\}$.

\section{Stokes' theorem and classical electrodynamics}
An alert reader might have encountered a similar situation in the case of classical electrodynamics \cite{Griffiths_2023} with the radial vector field of the following form
\begin{equation}\label{coulomb}
    V^i=\frac{1}{r^2} \delta^{i r}, \quad i=\{1,2,3\}.
\end{equation}
When we evaluate the divergence of this vector with respect to spherical coordinates 
\begin{equation}
    \text{div} V = \frac{1}{\sqrt{|g|}}\partial_i \left( \sqrt{|g|} V^i \right),
\end{equation}
in Euclidean space $\mathbb{R}^3$ endowed with the spherical metric
\begin{equation}
    ds^2=dr^2+r^2 d\Omega^2_{(2)}, 
\end{equation}
we quickly find out that it vanishes
\begin{equation}\label{divergencecoulomb}
    \text{div} V = \frac{1}{r^2} \partial_r (r^2 \frac{1}{r^2}) = \frac{1}{r^2} \partial_r (1) = 0. 
\end{equation}
Let us now consider the volume integral of this divergence term over a sphere of radius $r_c$, centered at the origin
\begin{equation}
    \int_{M}^{}d^3x \sqrt{|g|} \text{div} V = \int_{0}^{r_c} dr \int_{S^2}^{} d\Omega\, r^2\, \text{div} V = 0.
\end{equation}
On the other hand, if we make use of the Gauss's law\footnote{In this line of reasoning, we may consider Stokes's theorem and Gauss's law (which is essentially its incarnation) interchangeably.} and transform the volume integral into a surface integral
\begin{equation}
     \int_{\mathcal{M}}^{}d^3x \sqrt{|g|} \text{div} V =  \int_{\mathcal{M}}^{}d^3x \partial_i\left(\sqrt{|g|}V^i \right) = \int_{\partial \mathcal{M}}^{}d^2y \sqrt{|\gamma|} n_i V^i ,
\end{equation}
with $\sqrt{|\gamma|}=r_c^2 \sin{\theta}$ and the unit vector $n_i=\delta_{i r}$ on the boundary $r=r_c$, we get a non-zero result
\begin{equation}\label{coulombsurf}
    \int_{\partial \mathcal{M}}^{}d^2y \sqrt{|\gamma|} n_i V^i = \int_{S^2}^{}d\Omega \left( r_c^2 \frac{1}{r_c^2} \right)= 4 \pi .
\end{equation}
The source of this discrepancy is the point $r=0$, where the considered field \eqref{coulomb} diverges. Notice that the surface integral \eqref{coulombsurf} is independent of the radius $r_c$. Therefore, we ought to expect the result of $4 \pi$ to hold for a sphere of any radius, no matter how small. At the same time, the vanishing of eq. \eqref{divergencecoulomb} holds firmly everywhere, except the origin. Thus, all of the contribution must come from this singular point $r=0$. The resolution ought be also familiar. In the distributional sense, we define the divergence of the vector field to be
\begin{equation}
    \text{div} V = 4 \pi \delta^{(3)}(x),
\end{equation}
with the help of Dirac delta function $\delta(x)$ representing a point-like particle.

The very same steps are taken in the case of classical electrodynamics when we are interested in the total charge Q \cite{Carroll:2004st}. Indeed, if we recall the conservation law
\begin{equation}
    \nablac_\mu J^\mu = 0,
\end{equation}
where $J^\mu$ is the conserved current 4-vector, with the 0-th component $J^0=\rho$ representing the charge density - we may then define the total charge $Q_\Sigma$ passing through the hypersurface $\Sigma$ of a constant time as
\begin{equation}\label{totalcharge}
    Q_\Sigma=-\int_{\Sigma}^{}d^3y \, \sqrt{|\gamma |} \sigma_\mu J^\mu,
\end{equation}
with $\Sigma$ understood as some compact patch on the spacetime, i.e. not a boundary of any region, and $\sigma^\mu$ is the time-like normal vector to $\Sigma$. Next we make use of the Maxwell's equations in the following form
\begin{equation}
    \nablac_\mu F^{\nu \mu} = J^\mu,
\end{equation}
with tensor $F^{\mu \nu}$ constituting the field strength of electromagnetism. Substituting for the conserved 4-current in eq. \eqref{totalcharge}
\begin{equation}\label{volcharge}
    Q=-\int_{\Sigma}^{}d^3y \sqrt{| \gamma |} \sigma_\nu F^{\mu \nu}, 
\end{equation}
and employing the Stokes's theorem, we get the following equation for the total charge Q as
\begin{equation}
    Q=-\int_{\partial \Sigma}^{}d^2z \sqrt{| \gamma |}_{\partial \Sigma}\, n_\mu \sigma_\nu F^{\mu \nu},
\end{equation}
with $z^A,\, A=\{1,2\}$ being the coordinates on $\partial \Sigma$, $\gamma_{A B}|_{\partial \Sigma}$ is the induced metric and $n^\mu$ is the unit normal to $\partial \Sigma$. The electric field of a charge $q$ is related to the non-vanishing components of the field strength $F^{\mu \nu}$ as
\begin{equation}
    F^{t r}=-F^{r t}=E^r = \frac{q}{4 \pi r^2}.
\end{equation}
The normal vectors are this time set up as 
\begin{equation}
    \sigma^\mu=\delta^{\mu t}, \quad n^\mu=\delta^{\mu r},
\end{equation}
so that 
\begin{equation}
    n_\mu \sigma_\nu F^{\mu \nu} = - E^r = -\frac{q}{4 \pi r^2}.
\end{equation}
Lastly, we take $\partial \Sigma = S^2$ of some constant radius $r=r_c$ such that
\begin{equation}
    \sqrt{| \gamma |}_{\partial \Sigma}= r_c^2 \sin{\theta}. 
\end{equation}
All in all, we get the following expression for the total charge Q at the spatial infinity $r_c \rightarrow \infty$
\begin{equation}
    Q=-\int_{S^2}^{}d\Omega\, r^2 \left(-\frac{q}{4 \pi r^2} \right) = q.
\end{equation}
To obtain a matching result in the case when we interested in the volume integral \eqref{volcharge}, we but need to identify 
\begin{equation}
    \sigma_\mu \nablac_\nu F^{\mu \nu} = \frac{q}{4\pi} \delta^{(3)}(x),
\end{equation}
which represents a point-like particle of charge $q$ located at the origin.

\section{Canonical frame and the Dirac delta function}
If we recall the section~\ref{Canframe} on the canonical frame, we have encountered a similar situation as in the example above, when the teleparallel action vanished everywhere
\[
 S_{\text{TG}}=\int^{}_{} \frac{h}{2 \kappa} T = 0.
\]
Nonetheless, the Stokes' transformed boundary term remained non-trivial
\[
    S_{\text{bound}}=-\frac{1}{\kappa}\int_{r_c \rightarrow \infty}^{} d^3y \epsilon  \sqrt{|\gamma|} n_r  T^r |_{r=r_c}=\int_{0}^{\beta}d\tau\, \left( -\frac{M}{2} \right).
\]
Hence the logical conclusion is to amend this issue analogously as we did in the case of electrodynamics. We firstly inspect the singular behaviour of the vector torsion in \eqref{cfvector}
\begin{equation}
    T^r = - T^t = \frac{M}{r^2},
\end{equation}
which is well-defined on $\mathbb{R} ^{1,3}\setminus \{ 0\}$. Therefore, in order for us to extend this to the whole $\mathbb{R} ^{1,3}$, and at the same time ensure we get a non-zero result in the case of a volume integral, it is only natural to insist on following
\begin{equation}
    T=-\kappa M \delta^{(3)}(x) \equiv -\kappa M \left( \frac{1}{h}\, \delta(r)\, \delta(\theta)\, \delta (\phi)\right),
\end{equation}
where we have chosen the appropriate normalisation $-\kappa M$ based on the desired result of $-\frac{M}{2}$, and $\frac{1}{h}$ ensures the Dirac delta is normalised to 1 in the spherical coordinate system\footnote{For more details regarding the distributional nature of the black hole, see \cite{Balasin:1993fn}}. The teleparallel gravitational action for the exterior solution of the Schwarzschild black hole, when considered in the EF coordinates (canonical frame), is thus
\begin{equation}
     S_{\text{TG}}=\int^{}_{} \frac{h}{2 \kappa} T = -\frac{M}{2}\int^{}_{} dt \int^{\infty}_{0} dr\, \delta(r) \int^{\pi}_{0} d\theta\, \delta(\theta) \int^{2\pi}_{0}d\phi\, \delta(\phi)=\int^{}_{} dt\, \left(-\frac{M}{2}\right).
\end{equation}
At the same time, if we recall the interior solution for the canonical frame in the sense of the boundary term \eqref{cfintsol}, we obtained a trivial result despite the other approaches correctly predicting the value of $2M$. The reason for the vanishing prediction ought to be clear after the following rephrasing of the integral as
\[ S_{\text{bound}}=-\frac{1}{2}\int_{}^{} dt\, r_c^2  \left[  \frac{M}{r_c^2}\right]^{r_c=2M}_{r_c=\epsilon}=\]
\begin{equation}
=\int_{}^{}dt\, \left[ -\frac{M}{2} \right]^{r_c=2M}_{r_c=\epsilon}=0, 
\end{equation}
with the lower bound given as some non-zero, infinitesimal parameter $\epsilon > 0$ due to $ T^r =\frac{M}{r^2} \in \mathbb{R} ^{1,3}\setminus \{ 0\} $, i.e. the two boundaries cover the region outside of the ''point-charge" at the origin. Subtracting such two constant boundary terms inevitably leads to a trivial prediction.

Things get even weirder when we try to apply the same logic to the case of the proper frame and the vector torsion \eqref{vectorschw}. What at first seems like an innocent vector field (although singular at the origin $r=0$)
\[
    T^r=\frac{3M+2\left(-1+\sqrt{1-\frac{2M}{r}}\right)r}{r^2},
\]
it turns out the there is a discontinuity at the event horizon $r=2M$, illustrated in the following graph

\begin{figure}[!ht]
\centering
\includegraphics[width=.6\linewidth]{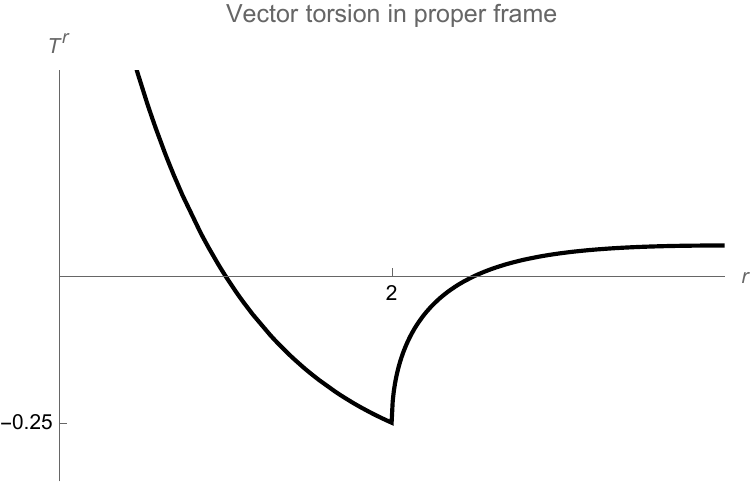}
\captionof{figure}{Graph of the real part of the only non-vanishing component of the vector torsion $T^\mu$ from eq. \eqref{vectorschw}.}
\end{figure}

Besides invalidating the use of Stokes' theorem, this discontinuity presents further difficulties when we try to modify the teleparallel action \eqref{tgrenaction} by the means of Dirac delta functions. At the same time, this demonstrates that the issue of the black hole action regularisation remains an open-ended question worth of a more refined analysis. We hope to delve deeper into this matter in our future research.  

\chapter*{Conclusion}
\addcontentsline{toc}{chapter}{Conclusion}
    \markboth{CONCLUSION}{CONCLUSION} 

To recall the GHY term approach, we had to firstly deal with the coordinate singularity located at the event horizon, which essentially split apart the internal and external solutions of the Schwarzschild black hole. Focusing on the exterior, we then analytically continued to the Euclidean version of the Schwarzschild solution by the means of Wick's rotation, all while losing any ability to describe the spacetime in the interior. In turn, we had to employ the GHY term for the sole boundary present - the one sent to infinity - and subtract the regulating term for the asymptotically flat background.

Then, we have explored the Schwarzschild solution in the context of teleparallel gravity by considering two frames: the diagonal and proper one. The subsequent regularisation of the action was realised by specifying the corresponding teleparallel spin connection by the method of reference tetrad. This concept of an observer located at infinity and its subsequent utilisation to find an appropriate spin connection is essentially based on the same idea as the background subtraction in general relativity. The overarching idea was to ensure that for the flat Minkowski spacetime, we recover trivial gravitational action. Additionally, by persisting in the telleparalel approach, we got a further insight into the behaviour of the Lagrangian density, with the possibility to analyse the contributions from both the interior and exterior of the black hole simultaneously; only to discover an inconsistency with the GH approach.

Unsettled by this discovery, we then ventured into the realm of generalised form of the Schwarzschild solution, the so-called Kerr-Schild form, in the hopes of utilising the concept of a canonical frame to discern between the considered models. But the free-falling nature of this frame, corresponding to the Schwarzschild metric in the EF coordinates, resulted in a vanishing teleparallel action. This was attributed to the failure of the reference tetrad method in providing us with a unique spin connection for a given frame.

We thus took an inspiration from the GH approach and proceeded analogously: we transformed the ensuing total divergence term to a boundary term by utilising the Stokes' theorem. With regards to the variation of the total action, this move perfectly mirrors the variational procedure of the EH action, where the role of the GHY term in erasing undesired second derivatives of the metric is replaced by a term proportional to vector torsion as is shown in \cite{Oshita:2017nhn,Erdmenger:2023hne}. The important distinction is that thanks to considering the teleparallel spin connection, this quantity requires no additional introduction of any regulating counter-terms. What is surprising is that in this way we obtained a result agreeing with GH approach with regards to the external solution, despite the total divergence term equaling zero before the employment of the Stokes' theorem.

We then further discussed the role of the Stokes' theorem and investigated the singular behaviour within the canonical frame. We argued that due to the free-falling nature of this frame, justified on the grounds of vanishing energy-momentum pseudotensor, we were unable to obtain unanimous results in both regions of the black hole. Albeit we were successful in fixing the volume integral by the means of Dirac delta function in analogy to classical electromagnetism in order to match the prediction for black hole action in the external region, the interior region remained trivial. This was attributed to the simple fact that the boundary term is constant, and as the interior solution is obtained as a difference of two boundary terms - one at the event horizon, the other infinitely close to the origin - the action consequently vanished altogether. 

On the other hand, the exterior solution was established in the sense of the Euclidean patch and the smoothing-out of the coordinate singularity at the event horizon. First of all, this resulted in only a single boundary being present, the one sent asymptotically to infinity. In this way, we were able to account for all of the contributions to the action inside the infinitely large two-sphere. But have we not already established that in the case of the canonical frame, the only contribution comes from the Dirac delta function residing at the origin? This apparently lead to some sort of logical inconsistency - as if the external solution, covering the region $r\in [2M, \infty)$ also accounted for the singularity inside the black hole. Nevertheless, if we recall that in the case of EF coordinates the metric is regular at the event horizon, there is no obstacle in the sense of a coordinate singularity that would prevent us to cover the whole Schwarzschild region. Such a line of reasoning fails if we were to try and apply it to the case of the proper frame. This opens up many new and interesting questions regarding the regularisation of the gravitational action.

Although we have been unable to completely explain away all of the inconsistencies of the regularised black hole action in the context of general relativity and teleparallel gravity, we have come to better understand where the issues arise and present the reader with a hint on the possible modifications needed to resolve them. Additionally, the development of alternative methods of regularisation and subsequent testing on more general black hole solutions seems like an intriguing continuation of this topic.

\bibliography{References}

\providecommand{\href}[2]{#2}\begingroup\raggedright\begin{thebibliography}{10}

\bibitem{Einstein:1905ve}
A.~Einstein, {\it {On the electrodynamics of moving bodies}},  Annalen Phys. {\bf 17} (1905) 891--921.

\bibitem{Einstein:1916vd}
A.~Einstein, {\it {The foundation of the general theory of relativity.}},  Annalen Phys. {\bf 49} (1916), no.~7 769--822.

\bibitem{Misner:1973prb}
C.~W. Misner, K.~S. Thorne, and J.~A. Wheeler, {\em {Gravitation}}.
\newblock W. H. Freeman, San Francisco, 1973.

\bibitem{EventHorizonTelescope:2019dse}
{\bf Event Horizon Telescope} Collaboration, K.~Akiyama et~al., {\it {First M87 Event Horizon Telescope Results. I. The Shadow of the Supermassive Black Hole}},  Astrophys. J. Lett. {\bf 875} (2019) L1, [\href{http://arxiv.org/abs/1906.11238}{{\tt arXiv:1906.11238}}].

\bibitem{Penrose:1969pc}
R.~Penrose, {\it {Gravitational collapse: The role of general relativity}},  Riv. Nuovo Cim. {\bf 1} (1969) 252--276.

\bibitem{Wald:1997wa}
R.~M. Wald, {\it {Gravitational collapse and cosmic censorship}},  in {\em {The Black Hole Trail}} (B.~Iyer, ed.), pp.~69--85.
\newblock Springer, 1997.
\newblock \href{http://arxiv.org/abs/gr-qc/9710068}{{\tt gr-qc/9710068}}.

\bibitem{Penrose:1999vj}
R.~Penrose, {\it {The question of cosmic censorship}},  J. Astrophys. Astron. {\bf 20} (1999) 233--248.

\bibitem{Bojowald:2010qpa}
M.~Bojowald, {\em {Canonical Gravity and Applications: Cosmology, Black Holes, and Quantum Gravity}}.
\newblock Cambridge University Press, 2010.

\bibitem{Rovelli:1997yv}
C.~Rovelli, {\it {Loop quantum gravity}},  Living Rev. Rel. (1998) [\href{http://arxiv.org/abs/gr-qc/9710008}{{\tt gr-qc/9710008}}].

\bibitem{Hawking:1979ig}
S.~W. Hawking, {\it {The path-integral approach to quantum gravity}},  in {\em {General Relativity: An Einstein Centenary Survey}} (S.~W. Hawking and W.~Israel, eds.).
\newblock Univ. Press, Cambridge, UK, 1979.

\bibitem{Hawking:1978jz}
S.~W. Hawking, {\it {Quantum Gravity and Path Integrals}},  Phys. Rev. D {\bf 18} (1978) 1747--1753.

\bibitem{Gibbons:1976ue}
G.~W. Gibbons and S.~W. Hawking, {\it {Action Integrals and Partition Functions in Quantum Gravity}},  Phys. Rev. D {\bf 15} (1977) 2752--2756.

\bibitem{York:1972sj}
J.~W. York, Jr., {\it {Role of conformal three geometry in the dynamics of gravitation}},  Phys. Rev. Lett. {\bf 28} (1972) 1082--1085.

\bibitem{Linder:2010py}
E.~V. Linder, {\it {Einstein's Other Gravity and the Acceleration of the Universe}},  Phys. Rev. D {\bf 81} (2010) 127301, [\href{http://arxiv.org/abs/1005.3039}{{\tt arXiv:1005.3039}}]. [Erratum: Phys.Rev.D 82, 109902 (2010)].

\bibitem{Einstein:1930xdd}
A.~Einstein, {\it {Auf die Riemann-Metrik und den Fern-Parallelismus gegr\"undete einheitliche Feldtheorie}},  Math. Ann. {\bf 102} (1930), no.~1 685--697.

\bibitem{moller1961conservation}
C.~Møller, {\it {Conservation Laws and Absolute Parallelism in General Relativity}},  {K. Dan. Vidensk. Selsk. Mat. Fys. Skr.} {\bf 1} (1961), no.~10 1--50.

\bibitem{Bahamonde:2021gfp}
S.~Bahamonde, K.~F. Dialektopoulos, C.~Escamilla-Rivera, G.~Farrugia, V.~Gakis, M.~Hendry, M.~Hohmann, J.~Levi~Said, J.~Mifsud, and E.~Di~Valentino, {\it {Teleparallel gravity: from theory to cosmology}},  Rept. Prog. Phys. {\bf 86} (2023), no.~2 026901, [\href{http://arxiv.org/abs/2106.13793}{{\tt arXiv:2106.13793}}].

\bibitem{Krssak:2018ywd}
M.~Kr\v{s}\v{s}\'ak, R.~J. van~den Hoogen, J.~G. Pereira, C.~G. B\"ohmer, and A.~A. Coley, {\it {Teleparallel theories of gravity: illuminating a fully invariant approach}},  Class. Quant. Grav. {\bf 36} (2019), no.~18 183001, [\href{http://arxiv.org/abs/1810.12932}{{\tt arXiv:1810.12932}}].

\bibitem{Aldrovandi:2013wha}
R.~Aldrovandi and J.~G. Pereira, {\em {Teleparallel Gravity}: {An Introduction}}.
\newblock Springer, 2013.

\bibitem{Brown:2015bva}
A.~R. Brown, D.~A. Roberts, L.~Susskind, B.~Swingle, and Y.~Zhao, {\it {Holographic Complexity Equals Bulk Action?}},  Phys. Rev. Lett. {\bf 116} (2016), no.~19 191301, [\href{http://arxiv.org/abs/1509.07876}{{\tt arXiv:1509.07876}}].

\bibitem{Brown:2015lvg}
A.~R. Brown, D.~A. Roberts, L.~Susskind, B.~Swingle, and Y.~Zhao, {\it {Complexity, action, and black holes}},  Phys. Rev. D {\bf 93} (2016), no.~8 086006, [\href{http://arxiv.org/abs/1512.04993}{{\tt arXiv:1512.04993}}].

\bibitem{Krssak:2023nrw}
M.~Kr\v{s}\v{s}\'ak, {\it {Bulk action growth for holographic complexity}},  Phys. Rev. D {\bf 109} (2024), no.~8 086002, [\href{http://arxiv.org/abs/2308.04354}{{\tt arXiv:2308.04354}}].

\bibitem{BeltranJimenez:2019bnx}
J.~Beltr\'an~Jim\'enez, L.~Heisenberg, and T.~S. Koivisto, {\it {The canonical frame of purified gravity}},  Int. J. Mod. Phys. D {\bf 28} (2019), no.~14 1944012, [\href{http://arxiv.org/abs/1903.12072}{{\tt arXiv:1903.12072}}].

\bibitem{Carroll:2004st}
S.~M. Carroll, {\em {Spacetime and Geometry}: {An Introduction to General Relativity}}.
\newblock Cambridge University Press, 2019.

\bibitem{Krssak:2024xeh}
M.~Kr\v{s}\v{s}\'ak, {\it {Teleparallel Gravity, Covariance and Their Geometrical Meaning}},  in {\em {Tribute to Ruben Aldrovandi}} ({F. Caruso, J.G. Pereira and A. Santoro}, ed.).
\newblock {Editora Livraria da Física}, São Paulo, 2024.
\newblock \href{http://arxiv.org/abs/2401.08106}{{\tt arXiv:2401.08106}}.

\bibitem{Fecko:2006zy}
M.~Fecko, {\em {Differential geometry and Lie groups for physicists}}.
\newblock Cambridge University Press, 2011.

\bibitem{Nakahara:2003nw}
M.~Nakahara, {\em {Geometry, topology and physics}}.
\newblock CRC Press, 2003.

\bibitem{Poisson:2009pwt}
E.~Poisson, {\em {A Relativist's Toolkit: The Mathematics of Black-Hole Mechanics}}.
\newblock Cambridge University Press, 2009.

\bibitem{Frolov}
V.~Frolov and A.~Zelnikov, {\em Introduction to Black Hole Physics}.
\newblock Oxford University Press, 2012.

\bibitem{Ortín_2004}
T.~Ortín, {\em Gravity and Strings}.
\newblock Cambridge University Press, 2004.

\bibitem{Krssak:2015lba}
M.~Kr\v{s}\v{s}\'ak, {\it {Holographic Renormalization in Teleparallel Gravity}},  Eur. Phys. J. C {\bf 77} (2017), no.~1 44, [\href{http://arxiv.org/abs/1510.06676}{{\tt arXiv:1510.06676}}].

\bibitem{Lucas:2009nq}
T.~G. Lucas, Y.~N. Obukhov, and J.~G. Pereira, {\it {Regularizing role of teleparallelism}},  Phys. Rev. D {\bf 80} (2009) 064043, [\href{http://arxiv.org/abs/0909.2418}{{\tt arXiv:0909.2418}}].

\bibitem{Krssak:2015rqa}
M.~Kr\v{s}\v{s}\'ak and J.~G. Pereira, {\it {Spin Connection and Renormalization of Teleparallel Action}},  Eur. Phys. J. C {\bf 75} (2015), no.~11 519, [\href{http://arxiv.org/abs/1504.07683}{{\tt arXiv:1504.07683}}].

\bibitem{Horowitz:1992jp}
G.~T. Horowitz, {\it {The dark side of string theory: Black holes and black strings.}},  in {\em String Theory and Quantum Gravity '92}, 1992.
\newblock \href{http://arxiv.org/abs/hep-th/9210119}{{\tt hep-th/9210119}}.

\bibitem{Ellis:1977pj}
G.~F.~R. Ellis and B.~G. Schmidt, {\it {Singular space-times}},  Gen. Rel. Grav. {\bf 8} (1977) 915--953.

\bibitem{Padmanabhan:2010zzb}
T.~Padmanabhan, {\em {Gravitation: Foundations and frontiers}}.
\newblock Cambridge University Press, 2014.

\bibitem{Frob:2021dmv}
M.~B. Fr\"ob, {\it {Kerr\textendash{}Schild metrics in teleparallel gravity}},  Eur. Phys. J. C {\bf 81} (2021), no.~8 766, [\href{http://arxiv.org/abs/2103.02620}{{\tt arXiv:2103.02620}}].

\bibitem{Oshita:2017nhn}
N.~Oshita and Y.-P. Wu, {\it {Role of spacetime boundaries in Einstein's other gravity}},  Phys. Rev. D {\bf 96} (2017), no.~4 044042, [\href{http://arxiv.org/abs/1705.10436}{{\tt arXiv:1705.10436}}].

\bibitem{BeltranJimenez:2018vdo}
J.~Beltr\'an~Jim\'enez, L.~Heisenberg, and T.~S. Koivisto, {\it {Teleparallel Palatini theories}},  JCAP {\bf 08} (2018) 039, [\href{http://arxiv.org/abs/1803.10185}{{\tt arXiv:1803.10185}}].

\bibitem{Emtsova:2021ehh}
E.~D. Emtsova, M.~Kr\v{s}\v{s}\'ak, A.~N. Petrov, and A.~V. Toporensky, {\it {On conserved quantities for the Schwarzschild black hole in teleparallel gravity}},  Eur. Phys. J. C {\bf 81} (2021), no.~8 743, [\href{http://arxiv.org/abs/2105.13312}{{\tt arXiv:2105.13312}}].

\bibitem{Griffiths_2023}
D.~J. Griffiths, {\em Introduction to Electrodynamics}.
\newblock Cambridge University Press, 5~ed., 2023.

\bibitem{Balasin:1993fn}
H.~Balasin and H.~Nachbagauer, {\it {The energy-momentum tensor of a black hole, or what curves the Schwarzschild geometry?}},  Class. Quant. Grav. {\bf 10} (1993) 2271, [\href{http://arxiv.org/abs/gr-qc/9305009}{{\tt gr-qc/9305009}}].

\bibitem{Erdmenger:2023hne}
J.~Erdmenger, B.~He\ss{}, I.~Matthaiakakis, and R.~Meyer, {\it {Gibbons-Hawking-York boundary terms and the generalized geometrical trinity of gravity}},  \href{http://arxiv.org/abs/2304.06752}{{\tt arXiv:2304.06752}}.

\end{thebibliography}\endgroup

\end{document}